\documentclass{article}
\usepackage[T1]{fontenc}
\usepackage[margin=25mm]{geometry}

\usepackage[ruled,linesnumbered,vlined]{algorithm2e}
\usepackage{graphicx}
\usepackage{listings}
\usepackage{xcolor}
\usepackage{subcaption} 
\usepackage{booktabs}
\usepackage{authblk}
\usepackage{mathtools} 
\usepackage[numbers]{natbib}
\usepackage{hyperref} 
\usepackage{amsthm} 
\usepackage{amssymb}
\newtheorem{theorem}{Theorem}
\newtheorem{definition}{Definition}
\newtheorem{example}{Example}

\definecolor{dkgreen}{rgb}{0,0.6,0}
\definecolor{gray}{rgb}{0.5,0.5,0.5}
\definecolor{mauve}{rgb}{0.58,0,0.82}
\SetCommentSty{mycommfont}
\newcommand{\osort}{\mathtt{OSort}}
\newcommand{\ocompact}{\mathtt{OCompact}}
\newcommand{\odistribute}{\mathtt{ODistribute}}
\newcommand{\opartition}{\mathtt{OPartition}}
\newcommand{\cmove}{\mathtt{CMove}}

\newcommand{\op}{\mathrm{op}}

\usepackage{tikz}
\usepackage{multirow}

\newcommand{\algname}{{\textsc{Jodes}}\xspace}
\newcommand{\dbucket}{{\textsc{DBucket}}\xspace}

\makeatletter
\DeclareRobustCommand\onedot{\futurelet\@let@token\@onedot}
\def\@onedot{\ifx\@let@token.\else.\null\fi\xspace}
\def\eg{\emph{e.g}\onedot} 
\def\ie{\emph{i.e}\onedot} 
 
\def\etc{\emph{etc}\onedot}

\def\resp{\emph{resp}\onedot}
\def\aka{\emph{aka}\onedot}
\makeatother

\newcommand{\ooplus}{\mathbin{\hat{\oplus}}}
\newcommand{\rev}[1]{#1}

\newif\iffull
\fulltrue  % Uncomment for the full version
\title{\algname: Efficient Oblivious Join in the Distributed Setting}

\author{Yilei Wang}
\author{Xiangdong Zeng}
\author{Sheng Wang}
\author{Feifei Li}
\affil{{\texttt{\{fengmi.wyl, zengxiangdong.zxd, sh.wang, lifeifei\}@alibaba-inc.com}}}
\affil{Alibaba Cloud}

\date{\vspace{-5ex}}

\begin{document}

\maketitle

\begin{abstract}
  Trusted execution environment (TEE) has provided an isolated and secure environment for building cloud-based analytic systems, but it still suffers from access pattern leakages caused by side-channel attacks. To better secure the data, computation inside TEE enclave should be made oblivious, which introduces significant overhead and severely slows down the computation. A natural way to speed up is to build the analytic system with multiple servers in the distributed setting. However, this setting raises a new security concern---the volumes of the transmissions among these servers can leak sensitive information to a network adversary. Existing works have designed specialized algorithms to address this concern, but their supports for equi-join, one of the most important but non-trivial database operators, are either inefficient, limited, or under a weak security assumption.

  In this paper, we present \algname, an efficient oblivious join algorithm in the distributed setting. \algname prevents the leakage on both the network and enclave sides, supports a general equi-join operation, and provides a high security level protection that only publicizes the input sizes and the output size. Meanwhile, it achieves both communication cost and computation cost asymptotically superior to existing algorithms. To demonstrate the practicality of \algname, we conduct experiments in the distributed setting comprising 16 servers. \rev{Empirical results show that \algname achieves up to a sixfold performance improvement over state-of-the-art join algorithms.}
\end{abstract}

\section{Introduction}

The provision of computation over encrypted data has become a crucial offering for cloud-based analytic system providers~\cite{10.1145/3318464.3386141,10.14778/3554821.3554826,10.14778/3476311.3476351,288576}.
The importance of encryption lies in its ability to ensure the confidentiality and privacy of data processed in the cloud, especially given the large amounts of sensitive and confidential information from enterprise customers. Data such as personal details, financial documents, trade secrets, and intellectual property require strong security measures to prevent unauthorized access or breaches. The deployment of encrypted data analytic systems~\cite{10.1145/3318464.3386141, 10.1145/2463676.2467797, 10.14778/3364324.3364331,opaque,10.1145/2043556.2043566, 8418608, 10.14778/3447689.3447705,10.14778/3554821.3554826,10.14778/3055330.3055334,Gribov2017StealthDBAS,10.14778/3583140.3583158}
guarantees the secure {computation and transmission of data},
ensuring that only authorized parties can access and decipher the information. This capability grants customers more authority over their data, enabling them to adhere to data governance protocols and address issues revolving around data privacy and sovereignty.

However, computation over encrypted data often incurs significant  overhead, typically resulting in several orders of magnitude slowdown. For instance, solutions based on secure multi-party computation~\cite{10.14778/3055330.3055334,263796} or homomorphic encryption~\cite{10.14778/3574245.3574248} can introduce at least three orders of magnitude
slowdown in computations. 
Meanwhile, the more practical approach builds encrypted analytic systems based on trusted execution environments (TEEs) like Intel SGX ~\cite{10.1145/3318464.3386141,10.14778/3364324.3364331,opaque,Gribov2017StealthDBAS,10.14778/3447689.3447705,10.14778/3554821.3554826,10.14778/3583140.3583158},
which is the focus of this paper. In this setup, cloud services operate within isolated \textit{enclaves} where confidential data is processed. Data must be decrypted only within the enclave for computation and re-encrypted before leaving the enclave, resulting in unavoidable encryption/decryption overhead. Besides, TEEs still suffer from an important vulnerability known as the \textit{access pattern leakage} caused by side-channel attacks~\cite{7163052,10.1145/2810103.2813695} where the host system can infer auxiliary information of encrypted data by monitoring memory accesses of the application~\cite{DBLP:conf/ndss/IslamKK12,10.1145/1037947.1024403}.
To mitigate this issue, computation in enclave should be designed as \textit{oblivious}~\cite{10.14778/3407790.3407814,10.1145/3514221.3517868}, ensuring that the access pattern of the computation is independent of the input. General techniques such as Oblivious RAM (ORAM)~\cite{10.1145/233551.233553} can be employed to transform non-oblivious algorithms into oblivious counterparts.
However, even the most practical ORAM scheme~\cite{10.1145/3177872} could significantly increase the running time by a factor of $O(\log^2 N)$ where $N$ is the input size, and the slowdown factor is between 90 and 450 according to the experiments of database join in~\cite{10.1145/3514221.3517868}. Such a high overhead prevents the encrypted analytic system from practicality.

There are two directions {to speed up the oblivious computation}:
(1) Design a specialized oblivious algorithm for each operator, which typically reduce the $O(\log^2 N)$ blow-up factor of ORAM to $O(\log N)$, \eg,  sorting~\cite{10.1145/1468075.1468121} and some database operations~\cite{DBLP:conf/icdt/ArasuK14,10.14778/3407790.3407814}; (2) Leverage the \textit{distributed setting}~\cite{10.1145/79173.79181,10.1145/1327452.1327492,10.5555/2228298.2228301}, which distributes intensive computation tasks to multiple servers, therefore {offsetting the overhead brought by obliviousness.}
Above two directions are orthogonal, and this paper studies their intersection point---designing efficient specialized oblivious algorithms in the distributed setting.

As firstly pointed out in~\cite{10.1145/2810103.2813695}, building encrypted analytic systems in the distributed setting raises new security concerns,
even though data are encrypted and processed inside enclave with obliviousness protection. Consider a network adversary that observes the communications between the servers. Although data are encrypted, their volumes are revealed,
which introduce the \textit{communication pattern leakage}, \ie, the number of elements transmitted between each pair of servers leaks information.
For example, a hash join operator will gather all the tuples that have the same join keys to the same server, which is determined by the hashing values. The network adversary could therefore infer some information about the distribution of join keys by analyzing the communication pattern.
To clearly distinguish the two types of leakages, we follow~\cite{mpc4mpc} and define a distributed algorithm  to be \textit{communication oblivious}, if its communication pattern is independent of the input (see Section~\ref{sec:oblivious} for a formal definition). The communication obliviousness defends against the network adversary. Accordingly, \textit{computation oblivious} is defined over a local computation, meaning the memory access pattern of the computation inside enclave is independent of the input, which defends against a memory adversary. All algorithms proposed in this paper are both communication and computation oblivious, which we simply refer to as \textit{oblivious}. For the case that only communication obliviousness is required, \eg, the servers are fully trusted, the system could implement the local computations in a natural way without the computation obliviousness requirement, which we ignore the details.

\subsection{Previous Work}
Opaque~\cite{opaque} and SODA~\cite{soda} are the only two prior works that present specialized oblivious distributed algorithms for join. Opaque starts by proposing the oblivious sorting algorithm based on column sort~\cite{10.1145/800057.808667}, and leverages it to implement oblivious filter, aggregate, and join. Opaque's join algorithm, however, is limited to only \textit{primary key join} (PK join), a special type of join where tuples from one of the input tables have unique join keys.
SODA~\cite{soda} considers column sort to be too expensive, so it proposes its own oblivious algorithms for filter, aggregate, and join without relying on oblivious sorting.
SODA's join supports a general binary equi-join, not limited to any special type of join. Nevertheless, it requires publicizing the \textit{degrees} (\ie, frequencies) of the most popular keys of the two tables, hence has a lower security level compared to other algorithms. 
In SODA's join algorithm, publicizing the maximum degrees is necessary for grouping tuples with the same keys to the same \textit{bins} to perform join. To achieve obliviousness, all bins should be padded to the same volume, which is computed from the maximum degrees.
We note that similar challenge also exists in the \textit{standalone setting} (\ie, a single machine is utilized to perform the join),
until the oblivious \textit{expansion} algorithm appears~\cite{10.14778/3407790.3407814}.
It then becomes the core building block of state-of-the-art oblivious standalone join algorithm~\cite{10.14778/3407790.3407814} without requiring any public degree information of the input tables. In this paper, we propose the first distributed expansion algorithm which also serves as one of the basic primitive of our join algorithm \algname.

\subsection{Our Contribution} 
The major contribution of this paper is \algname (Algorithm~\ref{alg:join}), an efficient \underline{J}oin algorithm that is \underline{O}blivious in the \underline{D}istribut\underline{E}d \underline{S}etting.
%, which is mainly based on our efficient oblivious expansion algorithm (Algorithm~\ref{alg:expansion}). 
Compared with previous works, it has the following advantages:
\begin{enumerate}
    \item Unlike Opaque's join, \algname supports a general binary equi-join operation, not limited to a primary key join. It also does not publicize any degree information as SODA's join, thus achieving higher security level than it (Section~\ref{sec:model}). 
    \item \algname is the first oblivious distributed join algorithm that achieves communication cost linear to only input size and output size. It also has computation cost asymptotically better than all existing works (Table~\ref{tab:compare}).
    \item Our experiments demonstrate that \rev{\algname outperforms all baselines}. For example, it can finish a join that outputs $1.9\times 10^8$ rows in 86s using 16 servers, which is only 1/6 of time taken by the the state-of-the-art distributed or standalone join algorithm (Section~\ref{sec:exp}).
\end{enumerate}

Apart from the expansion primitive, \algname also takes shuffle, sorting, and PK join as primitives. For sorting, we simply adopt the column sort in Opaque, while for shuffle and PK join, we have dedicatedly design faster oblivious algorithms for them (Section~\ref{sec:basic}). 
Experiments show that our algorithms for these operators improve the baselines by at least 60\% (Section~\ref{sec:exp}).
These operators are basic and instrumental, and our improvements should be of independent interest to the community of oblivious query processing beyond the scope of \algname itself. Any future refinement to the sorting operator can also enhance the performance of \algname.

\section{Preliminary}\label{sec:model}
The frequently used notations are summarized in Table~\ref{tab:notation}. 

\begin{table}
    \caption{Notations used in the paper}\label{tab:notation}
    \centering
    \begin{tabular}{ll} \toprule
    \textbf{Notation} & \textbf{Meaning} \\\midrule
    $p$ & Number of servers\\ 
    $[p]$ & The set $\{1,2,\dots,p\}$ \\ 
    $N$ (\resp $M$) & Number of input (\resp output) elements/tuples\\ 
    $n$ (\resp $m$) & $N/p$ (\resp $M/p$) \\
    $\sigma$ & Security parameter \\ 
    $X[i]$ & The part of elements/tuples in $X$ that are located on the $i$-th server \\ 
    $n_i$ & $|X[i]|$, num of elements/tuples on the $i$-th server \\ 
    $Y_{ij}$ & The subset of elements in $X_i$ that will be sent to the $j$-th server through communication  \\ 
    $U_i$ & The target size for padding of $Y_{ij}$ for any $j\in[p]$ \\\bottomrule
    \end{tabular}
    % \vspace{-1em}
\end{table}

\subsection{Distributed Setting}
In the \textit{distributed setting}, there are $p$ servers that work collaboratively to execute a computational task as dictated by a specific algorithm, which we refer to as a \textit{distributed algorithm}. Each server holds a portion of the input, which consists of $N$ elements almost equally distributed among them; the $i$-th server possesses a subset with $n_i=\Theta(N/p)$ elements, constituting its initial local dataset. The values $N$, $p$, and $\{n_i\}_{i=1}^p$ are public to all servers.
The servers complete the task over several \textit{rounds}. In each round, the $i$-th server processes its local dataset $X_i$ and generates output in the form  $Y_{i1},\dots,Y_{ip}$, for $i\in[p]$. This marks the computation phase for this round.
Following this, the servers enter the communication phase where all pairs of servers exchange data over a complete network. Specifically, the $i$-th server transmits the dataset $Y_{ij}$ to the $j$-th server. For any given server, the collective data received from all other servers are amalgamated to form its updated local dataset. This updated dataset is then either utilized in the subsequent round of computation, integrated into the input for the succeeding task, or emitted as the output.

To differentiate from the distributed setting, we employ the term \textit{standalone setting} to refer to the standard setting when there is only one server. An algorithm that is designed for the standalone setting is called a \textit{standalone algorithm}, which involves local computation on the server without any network communication.

\subsection{Encrypted Analytic System}
We focus on the cloud-based encrypted analytic system. The data on the servers is uploaded by one or more data owners in the encrypted form. When a client (who could also be one of the data owners) submits an (authorized) query to the servers, they translate the query to a query plan, \ie, a series of database operations, and then execute the operations following the underlying algorithms. The output of the last operation, which is also the output of the query, is then sent to the client, also in the encrypted form. The client has the key and can decrypt the query result into plaintext. This system securely enables querying across data from different owners, which may distrust each other as well as the servers.

In the system, each server is equipped with TEE, the size of whose protected memory is large enough to hold elements located in this server during computation. This is a reasonable assumption, thanks to the second-generation Intel SGX that supports enclave size up to 128GB or even larger~\cite{10.1145/3533737.3535098}.
All data elements are safeguarded with encryption while residing outside of the enclave, \ie, elements are in plaintext form inside enclave and are in ciphertext form outside enclave. The encryption scheme should defend against chosen-plaintext attack so that ciphertexts consistently appear random to the adversary, regardless of whether the corresponding plaintexts are identical. The length of a ciphertext is linear to the length of its underlying plaintext, which is independent of the plaintext's actual value. For example, AES with GCM encryption mode is sufficient. The TEEs should hold all DEKs (data encryption keys) of the data owners and the client, so that the input and output can be correctly decrypted and encrypted respectively. A common DEK for intermediate computations is held by all the TEEs on the servers, allowing ciphertexts received from one server to be correctly decrypted in enclave of another server. In the descriptions of our algorithms, we refer to an element without explicitly distinguishing its form, as it is clear that a plaintext (\resp ciphertext) is always inside (\resp outside) enclave. 

\subsection{Security Definition}\label{sec:oblivious}
\paragraph{Communication oblivious}
In the distributed setting, for any (deterministic or randomized) algorithm $\mathcal{A}$ and any input $X$, let $S(X)=\{s_{ijk}\}$ be a sequence where $s_{ijk}$ is the size of messages that the $i$-th server sends to the $j$-th server in the $k$-th round. Note that if $\mathcal{A}$ is random, then each $s_{ijk}$ is a random variable. This sequence $S(X)$ is all information that a network adversary can observe, which we call the \textit{transcript} of $\mathcal{A}$. \rev{We then have the following definition:
\begin{definition}\label{def:comm}
    $\mathcal{A}$ is communication oblivious if there exists a probabilistic simulator $\mathrm{Sim}$ such that, for any input $X=(X_1,\dots,X_p)$ where $|X_i|=n_i$ for all $i\in[p]$, the simulator can generate the simulation $\bar{S}=\mathrm{Sim}(n_1,\dots,n_p)$ such that, there is no polynomial-time algorithm that can distinguish between the transcript $S(X)$ and the simulation $\bar{S}$ with success probability more than $1/2$.
\end{definition}
}
In other words, the transcript of any input is only dependent on the input sizes across the servers, which ensures the network adversary infers no information of the input from the transcript (despite the input sizes). Note that the input sizes are usually assumed to be public. One may need to further protect them by, for example, differential privacy~\cite{10.1561/0400000042}, which is out the scope of this paper.

\paragraph{Computation oblivious}
For any \rev{standalone} algorithm $\mathcal{A}$  with input $X$ performed \textit{locally} on any server, its memory access operations of computing $\mathcal{A}(X)$ can be expressed by the sequence $((\op_1, a_1, x_1), \dots, (\op_k, a_k, x_k))$, where each $\op_i$ represents either a read or a write operation. Specifically, a read operation retrieves the element located at the $a_i$-th position inside enclave memory, while a write operation updates the element at the $a_i$-th position to $x_i$. The \textit{memory access pattern} of $\mathcal{A}$ with input $X$ is defined as $A(X)\coloneq (a_1,\dots,a_k)$, which is all the information that a memory adversary can observe through TEE side-channel attacks. \rev{We define computation oblivious as below:
\begin{definition}\label{def:comp}
    $\mathcal{A}$ is computation oblivious if there exists a probabilistic simulator $\mathrm{Sim}$ such that, for any input $X$ with $|X|=n$, the simulator generates $\bar{A}=\mathrm{Sim}(n)$ such that no polynomial-time algorithm can distinguish between $A(X)$ and $\bar{A}$ with success probability more than $1/2$.
\end{definition}
}

\paragraph{Security parameter} 
Our algorithms may fail to return correct answer. We define $\sigma$ as the \textit{security parameter}, and the theoretical analysis ensures that the failure probabilities of our algorithms are all bounded by $2^{-\sigma}$. Note that the adversary could not observe any failure, but the potential reaction to the failure may leak information, \eg, the client may re-submit the same query \rev{when a wrong answer is detected}. Therefore, it is necessary to choose a sufficiently large $\sigma$ (say, $\sigma=40$) so that the failure becomes almost impossible.

\paragraph{Dummy} During the execution of an oblivious algorithm, both computation and communication may involve \textit{dummy} elements. These elements serve as placeholders to maintain the algorithm's obliviousness without realistic meaning, as opposed to \textit{real} elements.  For example, to achieve communication oblivious, the $i$-th server may pad messages with dummy elements to a public and data independent size $U_i$ before sending to the $j$-th server (Section~\ref{sec:old_shuffle}).
To implement dummy elements, one may assign a unique value to it, guaranteed not to occur within the actual data domain, or alternatively, append an additional attribute to each element, indicating it as either dummy or real. Regardless of the chosen implementation strategy, it is crucial that dummy elements remain imperceptible to the adversary that:
(1) the ciphertexts of dummy and real elements are indistinguishable,
and (2) the access patterns to dummy and real elements are indistinguishable.

\subsection{Cost Model}
In evaluating a distributed algorithm, we consider both the \textit{communication cost} and the \textit{computation cost}. Regarding theoretical analysis, we treat each element (or tuple, in cases where the input comprises sets of tables) as a basic unit, and define input size (\resp output size) as the total number of elements or tuples within the input (\resp output).
The communication cost is the total number of elements that are communicated across the servers during the execution of an algorithm. Algorithms in Opaque~\cite{opaque}, SODA~\cite{soda}, and this paper, have communication cost linear to the input and output size, so we do not hide the constant of the linear term for detail comparison. However, we do neglect the lower-order terms. For instance, the communication cost of our oblivious shuffle by key algorithm is $N + o(N)$, and we omit the $o(N)$ term in our discussions. Meanwhile, the computation cost of an algorithm is the total costs of all the rounds, where the cost of each round is defined as the maximum cost of local computations of all the servers in the round. Computation cost typically scales superlinearly, especially with computation obliviousness. We express computation costs using asymptotic notations. 

In the encrypted model, data are encrypted prior to communication and re-decrypted afterward. We incorporate these encryption and decryption costs into the communication cost since they are performed (and only performed) before and after communication, and their costs are also linear to the number of elements exchanged between servers. Specifically, the time incurred by communication costs is proportional to a combination of network bandwidth and the speeds of decryption and encryption, introducing a considerable constant factor to the overall runtime. In contrast, computation cost, while superlinear, typically has a smaller constant factor dependent on CPU clock speed, memory frequency and latency, \etc. Given the varying significance of each cost in different scenarios, we strive in this paper to minimize both to the greatest extent possible.

\paragraph{Parameter assumptions}  Commonly, the security $\sigma$ is set between 40 and 80 in the literature~\cite{8584398,10.1145/3448016.3452808}. Given this context, our study will focus on inputs with size $N$ that satisfies both $N=\omega(p^2\sigma)$ and $N=O(2^\sigma)$. This ensures that $N$ is not overly small (rendering the distributed setting superfluous)
nor excessively large (exceeding contemporary servers' processing capabilities).
In the join algorithm, the costs are also determined by the public output size $M$. We make similar assumptions to $M$, \ie, $M=\omega(p^2\sigma)$ and $M=O(2^\sigma)$. Note that if $M$ is too small, the number of servers $p$ may be correspondingly reduced (in either a logical or physical sense) subsequent to the join operation, or it might even revert to the standalone setting.  We leverage these assumptions to simplify complexity expressions throughout this paper, \eg, it is obvious that $\log N=\Theta(\log n)$ and $\log m=\Theta(\log M)=O(\log N)$, where $n=N/p$ and $m=M/p$. Note that these assumptions only affect the performance analysis; correctness and security of our algorithms remain intact.

\subsection{Output Padding}\label{sec:output_padding}
\iffull
The security definition in Section~\ref{sec:oblivious} assumes the input size $N$ is public but not the output size $M$ as it is data dependent. 
Therefore, an oblivious algorithm will always pad the output to the worst output size over all inputs with size $N$.  For most database operations (\eg, filter, aggregate, PK join), this does not influence the overall performance of the algorithm much as the worst output size is only $O(N)$. However, for the join operation, the worst output size is the product of the sizes of the two input tables. Moreover, the composition of joins brings the computation cost to exponential, leading to poor performance or even unavailability of the system. Therefore, researches on oblivious query processing choose to sacrifice some security for better performance. Specifically, they propose different padding schemes to the output size of each join operation,
including no padding~\cite{10.14778/3407790.3407814,DBLP:conf/icde/HanZFLL22,soda}, padding to the next power of two~\cite{10.1145/3514221.3517868}, padding by differential privacy~\cite{10.14778/3291264.3291274}, \etc.   A critical issue of these padding schemes is that the leakage can aggregate if operators compose: when a query plan involves multiple joins, all intermediate (padded) join sizes, instead of only the output size of the query, are leaked, which could breach the security requirement and raise privacy issues. Therefore, not only the query, but also the query plan (\ie, how the operators compose), should also be evaluated and authorized before execution. One exception is the acyclic query, in which with proper preprocessing, intermediate join sizes are always bounded by the final join size~\cite{DBLP:conf/icdt/ArasuK14}, hence the intermediate leakage can be remove by setting the padding size of each intermediate join equal to that of the final join.

We support all the padding schemes, including the perfect secure one that pads to the worst size, by introducing the \textit{output size bound}~\cite{10.1145/3517804.3524142}. For any function $f$, let $\mathcal{A}$ be a distributed algorithm that implements $f$. For example, $f$ is the join operation while $\mathcal{A}$ is a distributed join algorithm. For any input $X$, despite the public input sizes on the servers $\{n_1,\dots,n_p\}$, an output size bound $M$ is also published to $\mathcal{A}$, which guarantees that $|f(X)|\le M$. Note that the value $M$ could be data dependent, and may be obtained by executing another oblivious algorithm priori to $\mathcal{A}$. The communication and computation of $\mathcal{A}$ can then depend on both $\{n_i\}$ and $M$. The output of $\mathcal{A}$ is with size exactly $M$, which consists of $f(X)$ and a set of dummy elements.  \algname leverages the output size bound $M$ to avoid the worst quadratic case. The way to determine $M$ raises an interesting research topic and is orthogonal to the study of this paper.

\else
The security definition in Section~\ref{sec:oblivious} assumes the input size $N$ is public but not the output size $M$ as it is data dependent. \rev{However, the worst-case output size of join operations can grow exponentially, resulting in poor performance or even system unavailability. Consequently, researchers working on  oblivious query processing have chosen to sacrifice certain security measures to achieve better performance. Specifically, they have proposed various padding schemes to standardize the output size of each join operation. Detailed explanations can be found in the full version of the paper \cite{full}.}
% However, the worst output size of joins are exponential, leading to poor performance or even unavailability of the system. Therefore, researches on oblivious query processing choose to sacrifice some security for better performance. Specifically, they propose different padding schemes to the output size of each join operation. We provide the details in the full version of the paper 
\fi
\section{Baseline}\label{sec:basic}
In this section, we review \rev{state-of-the-art distributed algorithms for the operators utilized by our join algorithm, as well as the sole existing oblivious distributed join algorithm. Our enhancements to some of these algorithms are presented in Section~\ref{sec:our_alg}.}

\subsection{Computation Oblivious Primitives}\label{sec:primitive}
Below we introduce some oblivious primitives that will be used in the local computations of our algorithms. All primitives introduced in this section only run \textit{locally} in the standalone setting, \ie, they are standalone primitives, therefore in Section~\ref{sec:primitive} by ``oblivious'' we mean computation oblivious.
These primitives form the basic building blocks to achieve computational obliviousness.

We denote the oblivious computation of the assertion $c$ as a binary value
with $[c]$ (\eg, $[x < y]$ has value 1 if $x < y$ and 0 otherwise).
We employ the notation $\cmove(z,x,a)$ to denote an oblivious subroutine that conditionally assigns the value of $a$ to $x$ if $z$ equals 1; if $z$ is 0, $x$ remains unmodified. The concrete implementations of the above operations could be based on assembly instructions \cite{10.5555/3241094.3241143} or branchless XOR-based C code \cite{sp24}.
We use $\bot$ to represent a dummy element or tuple, which is utilized solely for the purpose of padding and is designed to exert no influence on the outcome of the computation.

\paragraph{Sorting $\osort$} The bitonic sort~\cite{10.1145/1468075.1468121} stands as the most favored oblivious sorting algorithm, celebrated for its simplicity and practicality. It accomplishes the sorting of $n$ elements in $O(n\log^2 n)$ time. While there exist oblivious sorting algorithms with lower asymptotic complexity~\cite{10.1145/800061.808726,10.1145/2591796.2591830,doi:10.1137/1.9781611976014.2}, they are either non fully oblivious (assuming a super-constant sized trusted memory without obliviousness requirement), or encumbered by impractically large constant factors (outpace bitonic sort only when the input size is exceedingly large, which is an uncommon circumstance in the distributed setting). In this paper, we use $\osort(X, K)$ to represent an oblivious sorting operation that sorts $X$ by key $K$. Note that we will also discuss oblivious sorting under in the distributed setting that globally sorts the data across the servers. For disambiguation, we always use $\osort$ to refer to the locally sorting in the standalone setting, while using ``sorting'' to refer to the globally sorting in the distributed setting. 

\paragraph{Compaction $\ocompact$}
The compaction operator takes an array $X$ of $n$ elements and a binary array $M$ of length $n$ as input. The positions of $M$ with a value of 1 indicate ``marked'' elements, while positions with a value of 0 indicate ``unmarked'' elements. The compaction operation rearranges the elements in $X$ such that all marked items are positioned before unmarked items.  We use $\ocompact(X, M)$ to represent an oblivious compaction operation. Although it can be realized by invoking $\osort(X, M)$, we use the specialized oblivious algorithm for compaction in~\cite{10.1145/3548606.3560603}, which has only $O(n\log n)$ cost and is highly practical.

\paragraph{Distribution $\odistribute$}
Let $X=(x_1,\dots,x_n)$ be an array of non-dummy elements, and $T=(t_1,\dots,t_n)$ is an array with distinct values such that $t_i\in[m]$ for all $i\in[n]$, where $m\ge n$ is a public parameter. The distribution operator $\odistribute$ will output an array with size $m$ such that each $x_i$ is located at the $t_i$-th position for $i\in[m]$, while non-occupied positions are filled with dummy elements. \citet{10.14778/3407790.3407814} proposed an oblivious distribution algorithm with $O(m\log m)$ cost under the constraint that elements in $T$ are in ascending order. Note that this constraint can be removed if we apply $\osort((X,T),T)$ in advance, and then the total cost of $\odistribute$ will be $O(n\log^2 n+m\log m)$. We use $\odistribute(X, T, m)$ to represent an oblivious distribution operation.

\paragraph{Partitioning $\opartition$}
Let $X=(x_1,\dots,x_n)$ be an array of elements, and $T=(t_1,\dots,t_n)$ is an array such that: (1) Each $t_i\in [p]$; 
(2) Let $X_j\coloneq \{i\in[n]\mid t_i=j\}$, then $|X_j|\le U$ for all $j\in[p]$. 
We define the partitioning operator $\opartition$ as taking $X$ and $T$ as input, and outputs $p$ sequences $\{Y_j\}_{i=p}$, where each sequence $Y_j$ consists of $X_j$ and $U-|X_j|$ dummy elements, while the orders of them can be arbitrary. In other words, $Y_j$ is obtained by padding $X_j$ to the size bound $U$. In the distributed setting, $\opartition$ is an important primitive for a server to reorganize its local data $X$ by their specified targets $T$, where each $t_i\in[p]$ is the designated server that $x_i$ should be sent to from this server. After $\opartition$, the $j$-th output sequence $Y_j$ will be sent to the $j$-th server. As the total size of the sequences is $pU$, the blow up factor of $\opartition$ due to padding is $pU/n$. In this paper, to avoid oversized padding, our algorithms will always ensure that $U=O(n/p)$, so that the blow up factor is no more than a constant.
SODA~\cite{soda} has proposed an oblivious algorithm for $\opartition$ (Algorithm~1 in \cite{soda}). The idea is to first $\osort$ elements $X$ by $T$, compute the global position where each element should go to, and then apply $\odistribute$ the elements to these positions. The complexity of their algorithm is hence $O(n\log^2 n)$. 

\subsection{Shuffle}\label{sec:old_shuffle}
We then return to the distributed setting. If any sequence $X$ in a distributed algorithms is physically distributed across the servers, we use $X[i]$ to denote the segment of $X$ located on the $i$-th server. 

The most basic operator is \textit{shuffle}. Note that in this paper, shuffle does not mean random permutation (a procedure that puts data items in a uniformly random order). Instead, it refers to the operator for re-distributing data across servers in the distributed setting, as adopted in distributed data analytics engines such as Apache Spark \cite{10.5555/2228298.2228301}.  The shuffle operator takes two sequences $X$ and $T$ as input, where each $x\in X$ corresponds to a $t_x\in T$ which specifies the target server that $x$ should be sent to. Note that $X$ and $T$ locates across the $p$ servers, but each $(x,t_x)$ pair is in the same server. After the shuffle operator, the $j$-th server will receive $\{x\in X\mid t_x=j\}$, \ie, all elements in $X$ with  target $j$. For communication obliviousness, the shuffle operator also takes public parameters $\{U_i\}_{i=1}^p$ as input, and then the set of elements the $i$-th server receives will be padded to size $U_i$ by some dummy elements. Apparently, the shuffle operator can be implemented based on \rev{SODA's} $\opartition$, as shown in Algorithm~\ref{alg:shuffle}. Given $U_i=(1+o(1))(n_i/p)$ for all $i$, the communication cost is $N$ and the computation cost of each server is $O(n\log^2 n)$. 

\begin{algorithm}
    \caption{Shuffle}
    \label{alg:shuffle}
    \KwIn{$X$, $T$, and public parameters $\{U_i\}_{i=1}^p$ }
    \For{$i\gets 1$ \KwTo $p$}{
        $(Y_1,\dots,Y_p)\gets\opartition(X[i], T[i], U_i)$\;
        Send $Y_j$ to the $j$-th server for each $j\in[p]$\;
    }
\end{algorithm}

We use ``shuffle $X$ by  $T$'' to represent a shuffle operator with input $X$ and target servers $T$. Despite the standard definition, the shuffle operator also has two  commonly used variants:
\paragraph{Random shuffle} Random shuffle is a powerful operator that effectively eliminates the imbalance of the input. In the random shuffle operator, all $T$ are randomly chosen from $[p]$ uniformly and independently, \ie, all elements will be randomly shuffled across the $p$ servers. Since each $t\in T$ is independent to the input, it could be safely publicized without breaching the obliviousness definition, hence no padding is required. Therefore, instead of calling $\opartition$, a random shuffle operator simply groups the elements with the same target server in a natural (non-oblivious) way (\eg, by a length-$p$ array of lists). The computation cost can therefore be reduced to $O(n)$. We use ``shuffle $X$ randomly'' to represent a random shuffle operator with input $X$.

\paragraph{Shuffle by key}  Assume each element is in the key-value form $x=(k,v)$ where $k$ is the key and $v$ is the value. The shuffle by key operator defines $t_x=h(k)$, where $h$ is a random oracle\footnote{A random oracle is an ideal function that maps distinct elements to independent and uniformly random outputs taken from its range. In practice we suggest using a cryptographic hash function such as \href{https://github.com/BLAKE3-team/BLAKE3/}{BLAKE3}.} that is public to all servers. The shuffle by key operator can gather elements with the same key across the servers to the same server for further computation. Since the targets of shuffle by key are data dependent, we should apply the $\opartition$ for obliviousness. The parameters $\{U_i\}$ are determined by Theorem~\ref{thm:shuffle_by_key}.
We use ``shuffle $X$ by key $K$'' to represent a shuffle by key operator with input $X=(K,V)$ and its key $K$.

\begin{theorem}\label{thm:shuffle_by_key}
Setting $U_i=(1+c_i)n_i/p$, if the keys of $X[i]$ are all distinct for any $i$, then the shuffle by key algorithm fails with probability at most $2^{-\sigma}$, where 
    $c_i=\sqrt{{2.08p(\sigma+2\log p)}/{n_i}}=o(1).$
\end{theorem}

\iffull
\begin{proof}
Define $s_{ij}$ as the number of elements in $X[i]$ with target $j$, \ie, $s_{ij}=|\{x\in X[i]\mid t_x=j\}|$. Since the keys of $X[i]$ are all distinct, $s_{ij}$ follows the binomial distribution with parameters $n$ and $1/p$. Therefore, by Chernoff bound \cite{10.5555/1076315}, 
\[\Pr\left[s_{ij}>U_i\right]=\Pr\left[s_{ij}>(1+c_i)\frac{n_i}{p}\right]\le \exp\left(-\frac{c_i^2n_i}{3p}\right)=\frac{2^{-\sigma}}{p^2}.\]
Note that the algorithm fails only if there exists some $i\in[p]$ and $j\in [p]$ such that $s_{ij}>U_i$. By union bound, the probability of this event is at most $2^{-\sigma}$.
\end{proof}
\else
In this paper, we omit all the proofs, which can be found in the full version of the paper \cite{full}.
\fi

\subsection{Sorting}
The sorting operator permutes the original input such that for each server $i$, its local data $X[i]$ is sorted, and for any two servers $i$ and $j$ where $i<j$, $x\le y$ for any $x\in X[i]$ and $y\in X[j]$. Please note that the sorting operator in this section is for the distributed setting, while $\osort$ in Section~\ref{sec:primitive} is for the standalone setting.
Opaque~\cite{opaque} uses column sort~\cite{10.1145/800057.808667}, which is naturally oblivious.\footnote{Opaque's sorting algorithm is not inherently computation oblivious; however, substituting its local sorts with $\osort$ straightforwardly makes it computation oblivious. }
Column sort requires four rounds of local sorting and communication, with the communication costs for these rounds being $N$, $N$, $N/2$, and $N/2$ respectively. Thus, the total communication cost sums up to $3N$.
A recent study introduces \dbucket \cite{sp24}, a distributed sorting algorithm. \dbucket adapts the bucket oblivious random permutation proposed in \cite{doi:10.1137/1.9781611976014.2} to the distributed setting. This is followed by a non-oblivious distribution sort \cite{10.1145/48529.48535} (\aka sample sort). The bucket random permutation leads to a communication cost of $2N$ due to padding, while the non-oblivious sort contributes an additional $N$. Consequently, the total communication cost for \dbucket is also $3N$. %Since \dbucket has the same theoretical result compared with column sort, 
We employ column sort in our experiments due to its simplicity.
% \footnote{The authors of \dbucket did not perform a comparative analysis with the column sort algorithm, nor did they make their code publicly available in \cite{sp24}.An assessment of the actual performance differences between these two algorithms remains elusive.}

\subsection{\rev{Prefix Sum and Suffix Sum}}
Let $\oplus$ be a binary associative operator. The prefix sum operator takes $(x_1,\dots,x_N)$ as input and outputs $(x_1,x_1\oplus x_2,\dots,x_1\oplus x_2\oplus\dots \oplus x_N)$. The common choices of $\oplus$ are $+$, $\max$, $\min$, \etc
If each $x_i$ is in the key-value pair form $x_i=(k_i, v_i)$, then $\oplus$ is usually defined as\footnote{To implement $\oplus$ in an oblivious way, one could run $\cmove([k_1=k_2],v_2,v_1\ooplus v_2)$ and then simply outputs $(k_2,v_2)$.} 
\[
(k_1,v_1)\oplus (k_2,v_2) =\begin{cases}
    (k_2,v_1\ooplus v_2)  & \text{if }k_1=k_2, \\
    (k_2,v_2) & \text{otherwise,}
  \end{cases}
\]
where $\ooplus$ is another binary associative operator that operates on the values.  For example, in Opaque~\cite{opaque}, the stage 2--3 of oblivious aggregate is essentially a prefix sum operator where the key is the set of grouping attributes and $\ooplus$ is the aggregate function, and the stage 2--3 of oblivious sort-merge join (PK join in our paper) is also equivalent to a prefix sum operator where the key is the set of join attributes and $\ooplus$ always returns the first input, \ie, $v_1\ooplus v_2=v_1$.

The distributed algorithm for prefix sum operator~\cite{10.1007/978-3-642-25591-5_39} has communication cost $O(p)$, which is negligible compared with other operators as $p\ll N$. The algorithm is quite simple. First, each server $i$ locally computes the prefix sum on its input $X[i]$. Let $Y[i]$ be the output and $y_i$ be the last element of $Y[i]$, which is equal to sum of elements in $X[i]$. Each server sends $y_i$ to the first server, who then locally computes the prefix sum of $\{y_i\}_{i=1}^p$. Denote $\{z_i\}_{i=1}^p$ to be the output. The first server sends each $z_i$ to the $i+1$-th server for all $i\in[p-1]$. Finally each server $i\ge 2$ adds the element $z_{i-1}$ it receives to all the elements in $Y[i]$, \ie, updates each $y\in Y[i]$ to $z_{i-1}\oplus y$. Then $(Y[i])_{i=1}^p$ is the prefix sum of $(X[i])_{i=1}^p$. 
% The algorithm is oblivious as the above process is independent of the input.

We will also need the \textit{suffix sum} operator, which takes the same input as prefix sum but outputs $(x_1\oplus x_2\oplus\dots \oplus x_N, x_2\oplus\dots \oplus x_N,\dots,x_{N-1}\oplus x_N,x_N)$. It is trivial to implement oblivious suffix sum algorithm as the prefix sum operator in a symmetric way with the same costs, and we omit the details.

\subsection{\rev{Join}}\label{sec:intro_join}
In database theory, a \textit{join}\footnote{We only consider natural join (\aka equi-join) in this paper.} operator takes two tables $R$ and $S$ as input, and outputs the combinations of tuples from $R$ and $S$ that have the same values on the joined attributes (\aka join key). Without loss of generality, assume $R=R(A,B)$ and $S=S(B,C)$ and let the join key be $B$, then the join result of $R$ and $S$ is 
\[R(A,B)\Join S(B,C)=\{(a,b,c)\mid (a,b)\in R\land(b,c)\in S\}.\]

\rev{In the rest of this paper, we denote $N_1=|R|$, $N_2=|S|$, and $M=|R\Join S|$.} We define $\alpha_1$, the maximum degrees of the join key on $R$, as $\alpha_1\coloneq \max_{b_0}|\{(a,b)\in R\mid b=b_0\}|$. Similarly, $\alpha_2\coloneq \max_{b_0}|\{(b,c)\in S\mid b=b_0\}|$. \rev{We also define the \textit{$\ell_\infty$-skewness} of a join with output size $M$ as $\phi=\alpha_1\alpha_2/M$.}

\paragraph{Comparison between oblivious joins}
The theoretical comparison between our oblivious join \algname with existing ones is summarized in Table~\ref{tab:compare}.  The {standalone} algorithm works by all servers sending data to the first server who then performs state-of-the-art local oblivious join~\cite{10.14778/3407790.3407814} in the standalone setting,  splits the join result to $p$ parts, and sends each part to the corresponding server. \textit{Cartesian join} first ignores the join conditions and computes the cartesian product of the two input tables~\cite{5710932,10.1145/2594538.2594558}, and then filters the output tuples that does not meet the join conditions out by oblivious filter~\cite{soda}. It is notable that \algname has computation cost $O(1/p)$ of the standalone algorithm. The speed up factor $\Theta(p)$ means that \algname has perfectly balanced the computation to the $p$ servers asymptotically.

{SODA}~\cite{soda} proposed the first specialized oblivious algorithm for a general join. In addition to the total input size $N$ and the output size $M$, SODA's join algorithm also reveals $\alpha_1,\alpha_2$, the maximum degrees of the join key of the two tables. 
The key idea of SODA's join algorithm is to first arrange all the various-sized join groups into a set of equally-sized bins (first level assignment), and then distribute the bins to servers in a load balanced manner (second level assignment). Thereafter, each server computes the local join based on its assigned bins. To achieve obliviousness, the local join at each server produces an output of size $M/p+\alpha_1\alpha_2$ padded with some dummy tuples. \rev{Optionally, the dummy tuples could be} ultimately removed by SODA's filter algorithm.  Note that in second level assignment, the granularity of the involved shuffle is bins, with numbers bounded by $O(N/(\alpha_1+\alpha_2))$. As a result, it implicitly assumes $N=\Omega((\alpha_1+\alpha_2)p^2\sigma)$ to avoid padding by a super-constant factor, which means that it is infeasible to set $\alpha_1,\alpha_2$ to the worst sizes $N_1,N_2$ to achieve the same security level as other algorithms, where $N_1$ and $N_2$ are the sizes of the two input tables respectively. In conclusion, \algname has both costs asymptotically strictly better than all existing algorithms, except that when $\alpha_1\alpha_2=O(M/p)$, \rev{\ie, $p\phi=O(1)$}, SODA's join has the same complexity with \algname. But in any case,  SODA's join provides a weaker security guarantee than \algname.

\begin{table}
    \caption{Comparisons between oblivious join algorithms. Computation costs are presented asymptotically.} \label{tab:compare}
    \centering
    \begin{tabular}{|c|c|c|} \hline
    \textbf{Algorithm} & \textbf{Communication} & \textbf{Computation} \\\hline
    Standalone  & $N+M$ & $p(n+m)\log^2 n$  \\ \hline
    Cartesian join & $N_1N_2$ & $p(n\log n)^2$  \\ \hline
    SODA~\cite{soda} & $4N+\rev{(p\phi+1) M}$ & $(n+\rev{(p\phi+1) m})\log^2 n$ \\ \hline
    \algname (Ours) & $7N+2M+\min(2M,Np)$ & $(n+m)\log^2 n$ \\\hline
    \end{tabular}
    % \vspace{-1em}
\end{table}

\subsubsection{Primary Key Join}
We consider a special type of join, \textit{primary key join} (PK join), which guarantees that the join key is the primary key of $S$, \ie, All tuples in $S$ have distinct values in $B$. With this constraint on $S$, PK join typically gains more efficient algorithm than general join.  Opaque~\cite{opaque} supports oblivious PK join following the idea of sort-merge join. It calls the oblivious sorting operator twice on the union of the two tables, and the communication and computation costs of their algorithm are $6N_1+6N_2$ and $O(n\log^2 n)$ respectively, where $n=(N_1+N_2)/p$. 
The oblivious join algorithm in SODA~\cite{soda} can also be applied to PK join: By the primary key constraint, $\alpha_2=1$ and $M\le N_1$, hence it has communication cost $5N_1+4N_2+p\alpha_1$ and computation cost $O((n+\alpha_1)\log^2 n)$.

\section{Design}\label{sec:our_alg}

\rev{We propose the design of our algorithms in this section.}

\subsection{\rev{Shuffle}}\label{sec:our_shuffle}

% The computation cost of the shuffle algorithm (Algorithm~\ref{alg:shuffle}) comes from $\opartition$.
\rev{The computation cost of the shuffle algorithm (Algorithm~\ref{alg:shuffle}) is primarily attributed to $\opartition$.} We note that $\opartition$ does not need elements in each output sequence to be sorted, hence employing $\osort$ on all elements is superfluous. We borrow the idea of quicksort and propose our oblivious algorithm for $\opartition$ 
\iffull (Algorithm~\ref{alg:partition}).
\else as follows.\footnote{\rev{The algorithmic description is in the full version of this paper \cite{full}.}}
\fi
At the high level, quicksort is a recursive algorithm that partitions data to several buckets, where any element in $i$-th bucket  is not larger than any element in the $j$-th bucket for any $i<j$. Then it applies the quicksort algorithm on each bucket recursively. In $\opartition$, this recursion can be early stopped as long as the bucket size is at most $U$, hence the number of recursion levels can be reduced from $\log n$ to $\log p$. 
\iffull The details of each recursion of $\opartition$ is shown in Algorithm~\ref{alg:partition_recursion}.\fi  For each level, we choose the middle point as the pivot\iffull (Line~\ref{line:mid_pivot})\fi, and partition the elements to two parts according to the pivot by using $\ocompact$: Move all elements $(x_i,t_i)$ with $z_i=1$ in front of other elements in an oblivious way\iffull (Line~\ref{line:compact})\fi. Such $z_i$ could be determined by a linear scan\iffull (Line~\ref{line:z1}--\ref{line:z2})\fi. Afterwards, we input each of the two parts \iffull to Algorithm~\ref{alg:partition_recursion}\fi recursively. The cost of $\ocompact$ in each level is $O(n\log n)$ and the number of levels is $\lceil \log p\rceil$, so the total cost is $O(n\log n\log p)$.
\rev{Hence our algorithm reduces the computation cost of the shuffle operator from SODA's $O(n\log^2 n)$ to $O(n\log n\log p)$.}

\iffull
\begin{algorithm}
    \caption{Partitioning $\opartition$}
    \label{alg:partition}
    \KwIn{$\{(x_i,t_i)\}_{i=1}^n$ and public parameter $U$ }
    \KwOut{$Y_1,\dots,Y_p$ with $|Y_i|=U$ for all $i\in[p]$}
    Define $x_i=\bot$ and $t_i=0$ for all $n<i\le Up$\;
    $(Y_{1},\dots,Y_{p})\gets$ run Algorithm~\ref{alg:partition_recursion} with input $\{(x_i,t_i)\}_{i=1}^{Up}$ and public parameters $0,p,U$\;
    \Return{$(Y_{1},\dots,Y_{p})$}\;
\end{algorithm}

\begin{algorithm}
    \caption{Partitioning $\opartition$ recursion}
    \label{alg:partition_recursion}
    \KwIn{$\{(x_i,t_i)\}_{i=1}^{Up}$ and public parameters $l,r,U$}
    \KwOut{Sets $(Y_{l+1},\dots,Y_{r})$ with $|Y_i|=U$ for all $i$}
    $l'\gets lU$\;
    $r'\gets rU$\;
    \If{$r-l=1$}{
        \Return{$\{x_i\}_{i=l'+1}^{r'}$}\;
    }
    $m\gets \lfloor (l+r)/2 \rfloor$\;\label{line:mid_pivot}
    $m'\gets mU$\;
    $c\gets 0$\;\label{line:z1}
    \For{$i\gets l'$ \KwTo $r'$}{
        $c\gets c+[0< t_i\le m]$\tcp*{Non-dummy element that moves to the left side}
    }
    \For{$i\gets l'$ \KwTo $r'$}{
        $z_i\gets [t_i=0\land c<m'-l']$\tcp*{Dummy element that moves to the left side}
        $c\gets c+z_i$\;
        $z_i\gets z_i\lor [0< t_i\le m]$\;
    }\label{line:z2}
    $\ocompact(\{(x_i,t_i)\}_{i=l'+1}^{r'},\{z_i\}_{i=l'+1}^{r'}$)\; \label{line:compact}
    $(Y_{l+1},\dots,Y_{m})\gets$ run this algorithm recursively with input $\{(x_i,t_i)\}_{i=1}^{Up}$ and public parameters $l,m,U$\;
    $(Y_{m+1},\dots,Y_{r})\gets$ run this algorithm recursively with input $\{(x_i,t_i)\}_{i=1}^{Up}$ and public parameters $m,r,U$\;
    \Return{$(Y_{l+1},\dots,Y_{r})$\;}
\end{algorithm}
\fi

\subsection{\rev{Primary Key Join}}\label{sec:our_pkjoin}
Below we present our oblivious PK join algorithm (Algorithm~\ref{alg:pkjoin}) with lower costs.
Our basic idea follows the aggregate algorithm in SODA~\cite{soda} that tuples with the same key should be shuffled to the same server so that they can be joined locally. The main issue of simply invoking the shuffle by key operator is that it requires the tuples of the input table are distinct on their keys (Theorem~\ref{thm:shuffle_by_key}), which holds for $S$ but not for $R$. To resolve this issue, for tuples in $R$ with the same key in each server, we choose one of them as the \textit{representative} and mark other tuples as \textit{inactive}. In the shuffle by key operator, the representatives are shuffled by the join key while the inactive tuples are shuffled to random target servers independently. Then we can join the representatives of $R$ with all tuples of $S$ locally in each server (Line~\ref{line:pkjoin_start}--\ref{line:pkjoin_end}). Taking the information from $S$, the representatives then go back to their original servers (Line~\ref{line:shuffle_back}) and distribute the data they receive from $S$ to those inactive tuples (Line~\ref{line:distribute_start}--\ref{line:distribute_end}).  Note that in Line~\ref{line:pk_join_shuffle}, representatives have distinct $B$ but with $Z=0$ while inactive tuples have distinct and nonzero $Z$, so they are all distinct on $(B,Z)$, hence shuffle by key operator can be applied. Also note that Line~\ref{line:shuffle_back} is essentially the reverse of the shuffle in Line~\ref{line:pk_join_shuffle}, so they should have the same padding size.
Our algorithm has communication and computation cost $2N_1+N_2$ and $O(n\log^2 n)$ respectively.

\begin{algorithm}
    \caption{Oblivious PK join}
    \label{alg:pkjoin}
    \KwIn{$R(A,B)$ and $S(B,C)$ where $B$ is the primary key of $S$}
    \KwOut{$V(A,B,C)=R\Join S$}
    Add column $Z$ to $R[i]$ with $Z\gets 0$\;
    \For{$i\gets 1$ \KwTo $p$}{
        $\osort(R[i],B)$\;
        Add column $I$ to $R[i]$ with $I\gets i$ \tcp*{Record the original server id}
        \For{$j\gets 2$ \KwTo $|R[i]|$}{
            $(t_{j-1},t_j)\gets$ the $(j-1,j)$-th tuple of $R[i]$\;
            $\cmove([t_j.B=t_{j-1}.B], t_j.Z, j)$ \tcp*{Set inactive tuples to distinct and positive $Z$}
        }
    }
    Shuffle $R$ by key $(B,Z)$ \label{line:pk_join_shuffle} \tcp*{Inactive tuples are randomly shuffled}
    Shuffle $S$ by key $(B,0)$\;
    Initialize table $V(A,B,C,I,Z)$\;
    \For{$i\gets 1$ \KwTo $p$}{
        $n_0\gets |R[i]|$\;\label{line:pkjoin_start}
        Add column $C$ to $R[i]$\;
        Add columns $A,I,Z$ to $S[i]$ with $Z\gets -1$\;
        $V[i]\gets R[i]\cup S[i]$\;
        $\osort(V[i],(B,Z))$\;
        \For{$j\gets 2$ \KwTo $|V[i]|$}{
            $(t_{j-1},t_j)\gets$ the $(j-1,j)$-th tuple of $V[i]$\;
            $c\gets [t_j.B=t_{j-1}.B\land t_j.Z=0]$\;
            $\cmove(c,t_j.C,t_{j-1}.C)$\;
        }
        $\ocompact(V[i],[Z\ge 0])$ \tcp*{Move tuples from $R$ to the front} 
        Truncate $V[i]$ to size $n_0$\;\label{line:pkjoin_end}
    }
    Shuffle $V=(V[1],\dots,V[p])$ by $V.I$ \tcp*{Shuffle tuples back} \label{line:shuffle_back}
    \For{$i\gets 1$ \KwTo $p$}{
        $\osort(V[i],(B,Z))$\;\label{line:distribute_start}
        \For{$j\gets 2$ \KwTo $|V[i]|$}{
            $(t_{j-1},t_j)\gets$ the $(j-1,j)$-th tuple of $V[i]$\;
            $\cmove([t_j.B=t_{j-1}.B],t_j.C,t_{j-1}.C)$\; \label{line:distribute_end}
        }
    }
    Remove columns $I,Z$ from $V$\;
    \Return{$V$\;}
\end{algorithm}

\begin{example}
    Consider the example as shown in Figure~\ref{fig:pkfk}, in which there are two servers S1 and S2. The representatives of $R$ in S1 and S2 are $(a_1,1)$, $(a_2,2)$ and $(a_2,1)$, $(a_1,2)$ respectively. All other tuples are deemed inactive and represented in gray. Step (a) is the shuffle by key operation, during which representatives in $R$ and all tuples in $S$ are shuffled by their $B$ values ($h(2)=h(3)=1$ and $h(1)=h(4)=2$), whereas the inactive tuples are randomly assigned to a server. Step (b) entails executing a local PK join on each server. Note that inactive tuples are excluded from this join and instead have their $C$ values designated as $\bot$ (dummy). In step (c), all tuples are shuffled back to their original server. For instance, the tuple $(a_3,2,\bot)$, which was initially $(a_3,2)$ on S1, is relocated back to S1. Finally, in step (d), the active tuples distribute their $C$ values to the inactive tuples, thus completing the PK join process.
    \begin{figure}
\resizebox{\linewidth}{!}{
\begin{tabular}{c}
    \phantom{$v$}
    \begin{tabular}{|c||c|c|}
    \hline
    & $R(A,B)$ &   $S(B,C)$ \\ \hline
    S1 & \begin{tabular}{@{}c@{}} 
        $(a_1,1)$, \\ 
        $(a_2,2)$, \\ 
        {\color{gray} $(a_3,2)$,} \\
        {\color{gray} $(a_5,2)$ }  
        \end{tabular}  & \begin{tabular}{@{}c@{}} 
                        $(1,c_1)$, \\ 
                        $(2,c_2)$
                        \end{tabular}  \\ \hline
    S2 & \begin{tabular}{@{}c@{}} 
        $(a_2,1)$, \\
        {\color{gray}$(a_3,1)$,} \\ 
        $(a_1,2)$, \\ 
        {\color{gray}$(a_4,2)$} 
        \end{tabular} & \begin{tabular}{@{}c@{}} 
            $(3,c_1)$, \\ 
            $(4,c_3)$
            \end{tabular} \\ \hline
    \end{tabular}
    $\xrightarrow{\text{(a)}} $
    \begin{tabular}{|c|c|}
        \hline
         $R(A,B)$ &   $S(B,C)$ \\ \hline
        \begin{tabular}{@{}c@{}} 
            $(a_1,2)$, \\
            $(a_2,2)$, \\ 
            {\color{gray} $(a_3,1)$,} \\
            {\color{gray} $(a_5,2)$} 
        \end{tabular}  & \begin{tabular}{@{}c@{}} 
                            $(2,c_2)$, \\
                            $(3,c_1)$ 
                            \end{tabular}  \\ \hline
        \begin{tabular}{@{}c@{}} 
            $(a_1,1)$, \\ 
            $(a_2,1)$, \\
            {\color{gray}$(a_3,2)$,} \\
            {\color{gray}$(a_4,2)$} 
        \end{tabular} & \begin{tabular}{@{}c@{}} 
                            $(1,c_1)$, \\ 
                            $(4,c_3)$
                            \end{tabular} \\ \hline
    \end{tabular}
% \bigskip\\
    $\xrightarrow{\text{(b)}} $
    \begin{tabular}{|c|}
    \hline
     $T(A,B,C)$  \\ \hline
     \begin{tabular}{@{}c@{}} 
        $(a_1,2,c_2)$, \\
        $(a_2,2,c_2)$, \\ 
        {\color{gray} $(a_3,1,\bot)$,} \\
        {\color{gray}$(a_5,2,\bot)$}  
        \end{tabular}  \\\hline
     \begin{tabular}{@{}c@{}} 
        $(a_1,1,c_1)$, \\ 
        $(a_2,1,c_1)$, \\
        {\color{gray} $(a_3,2,\bot)$,} \\
        {\color{gray}$(a_4,2,\bot)$}
        \end{tabular}  \\ \hline
    \end{tabular}
    $\xrightarrow{\text{(c)}} $
    \begin{tabular}{|c|}
    \hline
     $T(A,B,C)$  \\ \hline
    \begin{tabular}{@{}c@{}} 
        $(a_1,1,c_1)$, \\ 
        $(a_2,2,c_2)$, \\ 
        {\color{gray}$(a_3,2,\bot)$,} \\
        {\color{gray}$(a_5,2,\bot)$}  \\
    \end{tabular}  \\\hline
    \begin{tabular}{@{}c@{}} 
        $(a_2,1,c_1)$, \\
        {\color{gray}$(a_3,1,\bot)$,} \\
        $(a_1,2,c_2)$, \\
        {\color{gray}$(a_4,2,\bot)$} 
    \end{tabular}  \\ \hline
    \end{tabular}
    $\xrightarrow{\text{(d)}} $
    \begin{tabular}{|c|}
        \hline
         $T(A,B,C)$  \\ \hline
        \begin{tabular}{@{}c@{}} 
            $(a_1,1,c_1)$, \\ 
            $(a_2,2,c_2)$, \\ 
            $(a_3,2,c_2)$, \\
            $(a_5,2,c_2)$  
        \end{tabular}   \\\hline
        \begin{tabular}{@{}c@{}} 
            $(a_2,1,c_1)$, \\
            $(a_3,1,c_1)$, \\
            $(a_1,2,c_2)$, \\
            $(a_4,2,c_2)$ 
        \end{tabular}  \\ \hline
        \end{tabular}
\end{tabular}
}
\caption{PK join algorithm example}
\label{fig:pkfk}
\end{figure}
\end{example}

\subsection{Expansion}
\rev{Before presenting our join algorithm, we need to introduce the expansion operator first.} Given a public parameter $M$, the expansion operator takes two arrays $X=(x_1,\dots,x_N)$ and $D=(d_1,\dots,d_N)$ as input, where each $d_i$ is a non-negative integer and $d_\bot\coloneq M-\sum_{i=1}^N d_i\ge 0$.  The values $d_i$ indicates the number of repetitions that $x_i$ should appear in the output, \ie, the output is a length-$M$ array:
\[(\underbrace{x_1,\dots,x_1}_{d_1\text{ times}},\underbrace{x_2,\dots,x_2}_{d_2\text{ times}},\dots,\underbrace{x_N,\dots,x_N}_{d_N\text{ times}},\dots,\underbrace{\bot,\dots,\bot}_{d_\bot\text{ times}}).\]
Note that those $x_i$ with $d_i=0$ would not appear in the output. The expansion operator was initially proposed for database join in~\cite{DBLP:conf/icdt/ArasuK14} and the oblivious standalone algorithm is formally described in~\cite{10.14778/3407790.3407814}. 
In this section, we propose \rev{the first} oblivious algorithm for the expansion operator in the distributed setting. 
Each server holds $N/p$ elements of $X$ and $D$ as input, and will hold $M/p$ elements of $Y$ as output after computation.
Our algorithm is described in Algorithm~\ref{alg:expansion}, in which we (logically) organize the input as a table $R(X,D)$ and output as a table $S(X)$ for better readability. 
% It has communication and computation cost $N+\min(M,Np)$ and $O(n\log n\log p+m\log^2 m)$ respectively.

Our algorithm works in two steps. Assume the output array $\{y_i\}_{i=1}^M$ is initially a length-$M$ array filled with dummy elements, \ie, $y_i=\bot$ for all $i\in[M]$.
Note that the largest index of each $x_i$ appearing in the output array is supposed to be $l_i\coloneq \sum_{j=1}^i d_j$, except that those with $d_i=0$ would not appear. The first step is to set $y_{l_i}$ to $x_i$ for each $i\in[N]$ for those $d_i>0$, and the second step is to replace each dummy element with the first non-dummy element after it (if there is), which could be realized by a suffix sum operator by defining proper $\oplus$ (Line~\ref{line:prefix_sum} in Algorithm~\ref{alg:expansion}).

To achieve the first step obliviously, we first note that the array $\{l_i\}_{i=1}^N$ can be obtained by calling a prefix sum operator with input $(d_1,\dots,d_{N})$.
Since each server will hold $m$ elements of the output array, the $l_i$-th element in the output will be held by server $t_i=\lceil l_i/m \rceil$, which suggests we should shuffle each $x_i$ to the $t_i$-th server (if $d_i=0$, $x_i$ is simply ignored). We denote this shuffle \texttt{SF1}.  Since the target servers in \texttt{SF1} are data dependent, it needs padding to achieve obliviousness. We perform a random shuffle \texttt{SF0} before \texttt{SF1} to balance the data, so that  the padding size of \texttt{SF1} is bounded.

\begin{theorem}\label{thm:expansion}
Let \texttt{SF0} be the random shuffle and \texttt{SF1} be the shuffle following \texttt{SF0}. If we set $U_i=(1+c_i)n_i\cdot \min(m/N,1)$ in \texttt{SF1}, then Algorithm~\ref{alg:expansion} has communication cost $N+\min(M,Np)$ and computation cost $O(m\log n+\min(m,N)\log^2 n)$ with failure probability at most $2^{-\sigma}$, where $n_i$ is the number of tuples on the $i$-th server after \texttt{SF0}, and
    $c_i=\sqrt{2.08\max(N/m,1)(\sigma+2\log p)/n_i}=o(1).$
\end{theorem}

\iffull
\begin{proof}
    The communication cost of \texttt{SF0} and \texttt{SF1} are $N$ and $p\sum_{i=1}^p U_i=\min(M,Np)$ respectively, hence the total communication cost is $N+\min(M,Np)$. The computation cost is dominated by $\odistribute(R[i].X, R[i].P, m)$, where the input size is $\sum_{i=1}^p U_i=\min(m,N)$. Hence the computation cost is $O(m\log n+\min(m,N)\log^2 n)$. 

    Next we bound the failure probability. Let ${m}_i$ be the number of tuples the $i$-th server receives after \texttt{SF1}. First note that these tuples are chosen from the input $R$, hence ${m}_i\le N$. Then we note that after expansion, each server should hold exact $m=M/p$ tuples, hence  ${m}_i\le m$. It suffices to consider the worst case ${m}_i=\hat{m}\coloneq \min(m,N)$ for all $i\in[p]$.

    Let $s_{ij}$ be the number of tuples that should be sent to the $j$-th server in \texttt{SF1}. Since the tuples have been randomly shuffle, $s_{ij}$ follows the hypergeometric distribution with parameters $N,\hat{m},n_i$. By Chernoff bound, 
    \[\Pr\left[s_{ij}>(1+c_i)\frac{\hat{m} n_i}{N}\right]\le \exp\left(-\frac{c_i^2n_i \hat{m}}{3N}\right)=\frac{2^{-\sigma}}{p^2}.\]
    Summing this probability over all $i\in[p]$ and $j\in[p]$, we conclude the theorem.
\end{proof}

\fi

\begin{algorithm}
  \caption{Oblivious expansion}
  \label{alg:expansion}
  \KwIn{$R(X,D)$, and public parameter $M$ }
  \KwOut{$S(X)$ where $t.X$ appears $t.D$ times for any $t\in R$}
  Add columns $(L,T,P)$ to $R$\; 
  $R.L\gets$ the prefix sum of $R.D$ \tcp*{Target global position}
  \For{$i\gets 1$ \KwTo $p$}{
    \For{$j\gets 1$ \KwTo $|R[i]|$}{
      $t_j\gets$ the $j$-th tuple of $R[i]$\;
      $t_j.T\gets\lceil t_j.L/m\rceil$ \tcp*{Target server}
      $t_j.P\gets t_j.L-(t_j.T-1)m$ \tcp*{Target position in the target server}
      $\cmove ([t_j.D=0],t_j,\bot)$\;
    }
  }
  Shuffle $R$ randomly \tcp*{No padding}
  Shuffle $R$ by $R.T$ with padding sizes specified by Theorem~\ref{thm:expansion}\;
  Initialize table $S(X)$\;
  \For{$i\gets 1$ \KwTo $p$}{
    $S[i].X\gets\odistribute(R[i].X, R[i].P, m)$\;
  }
  Run suffix sum operator on $S$ with $\oplus$ defined as: $x_1\oplus x_2$ is $x_2$ if $x_1=\bot$, otherwise $x_1$\label{line:prefix_sum}\;
  \Return{$S$}\;
\end{algorithm}

\begin{example}
Consider the example in Figure~\ref{fig:expansion} with $p=3$, $M=18$, and  $d_\bot=2$. Each server will hold $m=M/p=6$ elements of the output. Our algorithm first computes the prefix sum of $(1,3,1,0,5,2,1,1,2)$ as $L$ in step (a), indicating that $x_i$ will lastly appear at the $l_i$-th location in the output array, except that $d$ will not appear. This in turn implies $x_i$ will lastly appear in the $p_i$-th location of the $t_i$-th server, with $T$ and $P$  locally computed in step (b). In step (c), we shuffle $R$ randomly,  then shuffle it with target servers of tuples specified by $T$ with proper padding. Afterwards, each server locally put each tuple $t$ at the $t.P$-th position in its server by $\odistribute$. The result is shown as the fourth table in this figure. Step (d) is a suffix sum operation as described in Line~\ref{line:prefix_sum} of Algorithm~\ref{alg:expansion} which finally yields the expansion result.
  \begin{figure}
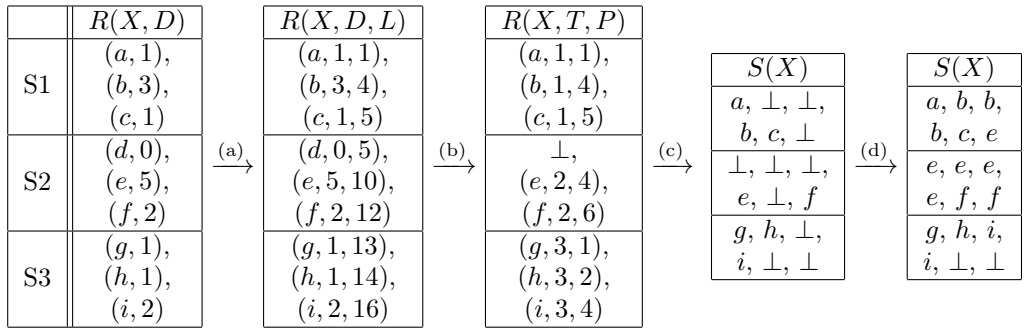

\centering
    \begin{tabular}{|c||c|}
    \hline
    & $R(X,D)$  \\ \hline
    S1 & \begin{tabular}{@{}c@{}} 
        $(a,1)$, \\ 
        $(b,3)$, \\ 
        $(c,1)$ 
        \end{tabular}    \\ \hline
    S2 & \begin{tabular}{@{}c@{}} 
        $(d,0)$, \\ 
        $(e,5)$, \\ 
        $(f,2)$ 
        \end{tabular}    \\ \hline
    S3 & \begin{tabular}{@{}c@{}} 
        $(g,1)$, \\ 
        $(h,1)$, \\ 
        $(i,2)$ 
        \end{tabular}    \\ \hline
    \end{tabular}
    $\xrightarrow{\text{(a)}} $
    \begin{tabular}{|c|}
        \hline
         $R(X,D,L)$  \\ \hline
        \begin{tabular}{@{}c@{}} 
            $(a,1,1)$, \\ 
            $(b,3,4)$, \\ 
            $(c,1,5)$ 
        \end{tabular}    \\ \hline
        \begin{tabular}{@{}c@{}} 
            $(d,0,5)$, \\ 
            $(e,5,10)$, \\ 
            $(f,2,12)$ 
        \end{tabular}    \\ \hline
        \begin{tabular}{@{}c@{}} 
            $(g,1,13)$, \\ 
            $(h,1,14)$, \\ 
            $(i,2,16)$ 
        \end{tabular}    \\ \hline
    \end{tabular}
    $\xrightarrow{\text{(b)}} $
    \begin{tabular}{|c|}
        \hline
         $R(X,T,P)$  \\ \hline
        \begin{tabular}{@{}c@{}} 
            $(a,1,1)$, \\ 
            $(b,1,4)$, \\ 
            $(c,1,5)$ 
        \end{tabular}    \\ \hline
        \begin{tabular}{@{}c@{}} 
            $\bot$, \\ 
            $(e,2,4)$, \\ 
            $(f,2,6)$ 
        \end{tabular}    \\ \hline
        \begin{tabular}{@{}c@{}} 
            $(g,3,1)$, \\ 
            $(h,3,2)$, \\ 
            $(i,3,4)$ 
        \end{tabular}    \\ \hline
    \end{tabular}
% \bigskip\\
    $\xrightarrow{\text{(c)}} $
    \rev{
    \begin{tabular}{|c|}
        \hline
         $S(X)$  \\ \hline
        \begin{tabular}{@{}c@{}} 
          $a$, $\bot$,  $\bot$,\\
          $b$,  $c$, $\bot$ 
        \end{tabular}   \\ \hline
        \begin{tabular}{@{}c@{}} 
          $\bot$, $\bot$,  $\bot$, \\
          $e$,  $\bot$, $f$
        \end{tabular} \\ \hline
        \begin{tabular}{@{}c@{}} 
          $g$, $h$, $\bot$, \\
          $i$, $\bot$, $\bot$
        \end{tabular} \\ \hline
    \end{tabular}
    $\xrightarrow{\text{(d)}} $
    \begin{tabular}{|c|}
        \hline
        $S(X)$  \\\hline
        \begin{tabular}{@{}c@{}} 
          $a$, $b$,  $b$,\\
          $b$,   $c$, $e$
        \end{tabular} \\\hline
         \begin{tabular}{@{}c@{}} 
          $e$, $e$,  $e$,\\
          $e$,  $f$, $f$ 
        \end{tabular} \\\hline
        \begin{tabular}{@{}c@{}} 
          $g$, $h$,  $i$, \\
          $i$, $\bot$, $\bot$
        \end{tabular} \\\hline
    \end{tabular}
    }
% \end{tabular}
\caption{Expansion algorithm example with $M=18$}
\label{fig:expansion}
\end{figure}
\end{example}

\subsection{Oblivious Join}
Now we are ready to present our oblivious join algorithm.
Note that our idea for PK join is not directly applicable to a generalized join operation, because for any $b$, there could be multiples tuples in both $R$ and $S$ with $B=b$. While it is still feasible to select any tuple from $R$ with $B = b$ as a representative, it is not known how to efficiently associate all corresponding tuples in $S$ with $B = b$ to this chosen representative.

We start by revisiting the state-of-the-art {standalone} oblivious join algorithm~\cite{10.14778/3407790.3407814}. The high level idea is based on the observation that for any $(a,b)\in R$, it appears $\deg_S(b)$ times in the join result, where $\deg_S(b)$ is the number of tuples in $S$ with $B=b$, which we call the \textit{degree} of $b$ in $S$. These degrees can be computed by combining sorting and prefix sum operators, and can be attached to the correct tuples in $R$ by a PK join operator. Then an expansion operator expands $R$ according to the degree of 
$S$, increasing the total size to $M$, the output table size. Note that this expansion requires $M$ as introduced in Section~\ref{sec:output_padding}.  These steps are then applied to $S$ symmetrically. Finally, it aligns the two expanded tables properly by an extra sorting on $S$ by join key and its \textit{alignment key}, which could be computed by the degrees of the two tables.
Note that the above algorithm is essentially the composition of the sorting, aggregate, PK join, and expansion operators. By instantiating these operators with our proposed distributed oblivious algorithms, it is transformed to the distributed version correctly.

Our oblivious join algorithm \algname is presented in Algorithm~\ref{alg:join}, and the subroutine that computes the degrees of the two tables are described in Algorithm~\ref{alg:join-sub}. Despite following the idea of the standalone oblivious join algorithm, we also optimize \algname in distributed setting by noting that the final alignment can be implemented without the sorting operator. Specifically, the original alignment key $L$ indicates the positions of the tuples in each group of $B$. We redefine $L$ so that it indicates the global positions, which are computed as in Line~\ref{line:computeL_start}--\ref{line:computeL_end}.   Instead of simply performing global sorting on $L$, we first compute the target servers $T$ of the tuples by the alignment key $L$ (Line~\ref{line:compute_dest}), shuffle the table by $T$, and then perform $\osort$ on the alignment key in each server. However, the target servers are data dependent, hence padding is required. Similar to our expansion algorithm, we perform a random shuffle in advance to balance the data, and setting padding size as stated in Theorem~\ref{thm:join} is adequate. 
The communication costs of the first 6 lines are $3N_1+3N_2,2N_1+N_2,0,N_1+\min(M,N_1p),N_1+2N_2,N_2+\min(M,N_2p)$ respectively, and the communication cost of the each of the two shuffles is $M$. Other operators involve only costs with low-order term.  Hence the total communication cost of \algname is $7(N_1+N_2)+\min(M,N_1p)+\min(M,N_2p)+2M\le 7N+2M+\min(2M, Np)$ where $N=N_1+N_2$.
The computation cost of \algname is dominated by $\osort$ before and after expansion, which is  $O((n+m)\log^2 n)$.

Algorithm~\ref{alg:join} assumes $M$ is a public parameter. If the padding scheme is ``no padding'', \ie, $M$ is the exact output size of the join as in SODA~\cite{soda}, then we can simply compute $M$ by summing all the degrees that $R$ receives after PK join, without the need of it being public. Specifically, insert ``$M\gets $ sum of ${R'}.D_S$'' after Line~\ref{line:insert_after}.

\begin{theorem}\label{thm:join}
Let \texttt{SF0} be the random shuffle (Line~\ref{line:join_random_shuffle}) and \texttt{SF1} be the other shuffle  (Line~\ref{line:join_shuffle}) in Algorithm~\ref{alg:expansion}. If we set $U_i=(1+c_i)n_i/p$ in \texttt{SF1}, then it fails with probability at most $2^{-\sigma}$, where $n_i$ is the number of tuples on the $i$-th server after \texttt{SF0}, and
    $c_i=\sqrt{{2.08p(\sigma+2\log p)}/{n_i}}=o(1).$
\end{theorem}

\iffull
\begin{proof}
    Note that the number of tuples the $i$-th server receives after \texttt{SF1} is exactly $m$ for any $i\in[p]$. 
    Let $s_{ij}$ be the number of tuples that should be sent to the $j$-th server in \texttt{SF1}. Since the tuples have been randomly shuffle, $s_{ij}$ follows the hypergeometric distribution with parameters $M,{m},n_i$. By Chernoff bound, 
    \[\Pr\left[s_{ij}>(1+c_i)\frac{n_i}{p}\right]\le \exp\left(-\frac{c_i^2n_i}{3p}\right)=\frac{2^{-\sigma}}{p^2}.\]
    Summing this probability over all $i\in[p]$ and $j\in[p]$, we conclude the theorem.
\end{proof}

\fi

\begin{algorithm}
  \caption{Oblivious join \algname}
  \label{alg:join}
  \KwIn{$R(A,B)$ and $S(B,C)$; public output size bound $M$}
  \KwOut{$V(A,B,C)=R(A,B)\Join S(B,C)$}
  Sort both $R$ and $S$ by $B$\;\label{line:preparation_start}
  ${R'}(A,B,D_R,D_S)\gets $ run Algorithm~\ref{alg:join-sub} with input $R,S$\;\label{line:insert_after}
  Remove column $D_R$ from $R'$\;
  ${S'}(B,C,D_R,D_S)\gets $ run Algorithm~\ref{alg:join-sub} with input $S,R$\;\label{line:preparation_end}
  $\bar{R}(A,B)\gets $ expansion with input $({R'}.A,{R'}.B)$, $({R'}.D_S)$ and $M$\;\label{line:expansion_start}
  $\bar{S}(B,C,D_R,D_S)\gets $ expansion with input $S'$, ${S'}.D_R$ and $M$\;\label{line:expansion_end}
  Add column $I,J,L,T$ to $\bar{S}$\; \label{line:computeL_start}
  $\bar{S}.I\gets$ prefix sum on key-value pair $(B,1)$\;
  $\bar{S}.J\gets$ prefix min on key-value pair $(B,[M])$\;
  \For{$i\gets 1$ \KwTo $p$}{
    \For{$t\in \bar{S}[i]$}{
        $q\gets t.I-1$\;
        $t.L\gets \lfloor q/t.D_R \rfloor + (q \bmod{t.D_R})\cdot t.D_S+t.J$\; \label{line:computeL_end}
        $t.T\gets \lceil t.L/m \rceil$\;\label{line:compute_dest}
    }
  }
  Shuffle $\bar{S}$ randomly\;\label{line:join_random_shuffle}
  Shuffle $\bar{S}$ by $S.T$ with padding size specified by Theorem~\ref{thm:join}\; \label{line:join_shuffle}
  Initialize table $V(A,B,C)$\;
  \For{$i\gets 1$ \KwTo $p$}{
    $\osort(\bar{S}[i],L)$\;
    \For{$j\gets 1$ \KwTo $m$}{
        $(t_{R},t_{S})\gets$ the $j$-th tuple of $(\bar{R}[i],\bar{S}[i])$\;
        Insert $(t_{R}.A,t_{R}.B,t_{S}.C)$ to $V[i]$\;
    }
  }
  \Return{$V$}\;\label{line:alignment_end}
  % \nonl\small\textsuperscript{1}In the comments,  ``cost'' refers to communication cost.
\end{algorithm}

\begin{algorithm}
  \caption{Compute degrees}
  \label{alg:join-sub}
  \SetKw{KwOr}{or}
  \KwIn{$R(A,B)$ and $S(B,C)$, both ordered by $B$ \;}
  \KwOut{$R(A,B,D_R,D_S)$}
  Add column $D_R$ to ${R}$\;
  ${R}.D_R\gets$ prefix sum on key-value pair $(B,1)$\;
  ${R}.D_R\gets$ suffix max on key-value pair $(B,D_R)$, \ie, $\oplus$ is defined as $x\oplus y = \max(x,y)$\;
  Add column $D_S$ to ${S}$\;
  $S.D_S\gets$ prefix sum on key-value pair $(B,1)$\;
  ${S}.D_S\gets$ suffix max on key-value pair $(B,D_S)$\;
  \For{$i\gets 1$ \KwTo $p$}{
    \For{$j\gets 2$ \KwTo $|S[i]|$}{
        $(t_{j-1},t_{j})\gets$ the $(j-1,j)$-th tuple of $S[i]$\;
        $\cmove([t_{j-1}.B=t_j.B],t_{j-1},\bot)$\tcp*{Remove duplicates}
    }
  }
  $R(A,B,D_R,D_S)\gets R(A,B,D_R)\Join S(B,C,D_S)$ \tcp*{PK join with communication cost $2|R|+|S|$}
  \Return{$R$}\;
\end{algorithm}

\begin{example}
  Consider the example shown in Figure~\ref{fig:join}, in which the output size bound $M$ is set to the true join size, \ie, no padding. The subroutine Algorithm~\ref{alg:join-sub} corresponds to step (a), which includes two sub-steps: (a1) computing the prefix sum on key-value pair $(B,1)$ to get $D_R$, and (a2) updating $D_R$ by suffix max and then obtaining $D_S$ by PK join. Step (b) is to apply the expansion operator on the degree of the other table. Besides, for $\bar{S}$, it also computes the alignment key $L$ and the target servers $T$. In step (c), we apply the two shuffle operators and then local $\osort$ so that $\bar{S}$ is ordered by $L$. The final step (d) is to combine $\bar{R}$ and $\bar{S}$ to get the join result $V$.
  {% \fontsize{8.5pt}{10pt}\selectfont
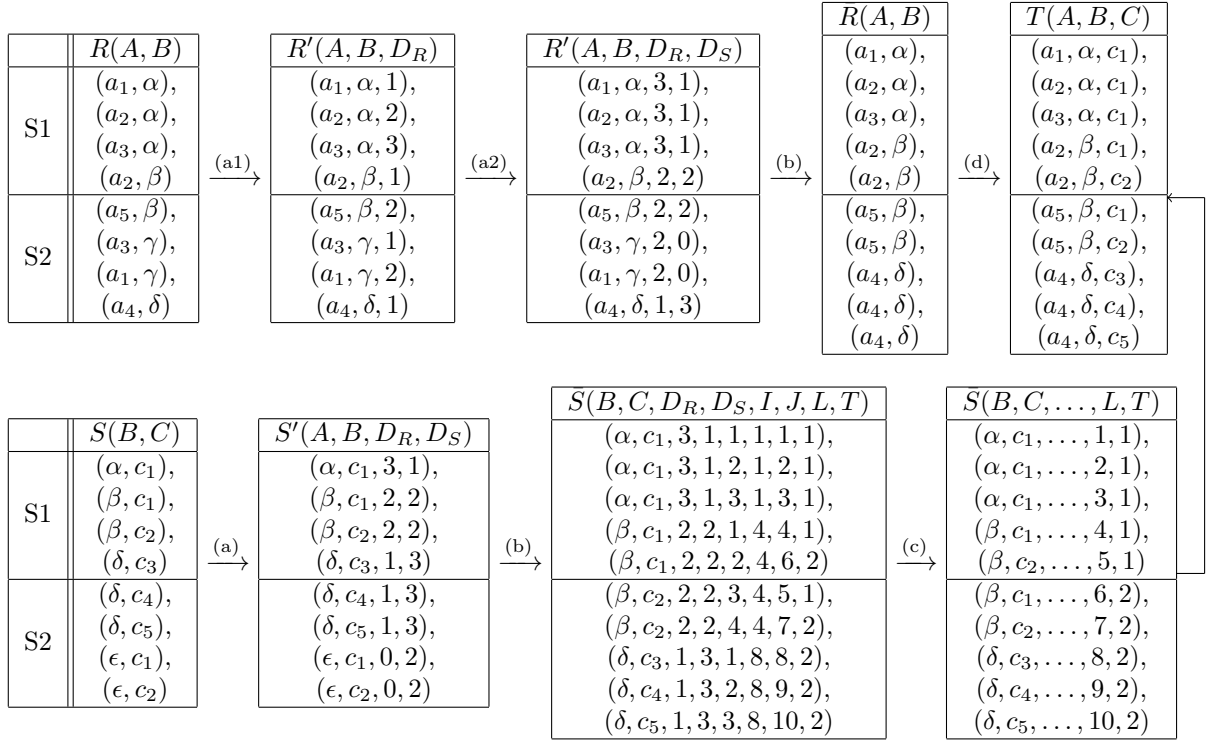
\begin{figure}
\centering
\begin{tabular}{l}
    \begin{tabular}{|c||c|}
    \hline
    & $R(A,B)$  \\ \hline
    S1 & \begin{tabular}{@{}c@{}} 
        $(a_1,\alpha)$, \\ 
        $(a_2,\alpha)$, \\
        $(a_3,\alpha)$, \\ 
        $(a_2,\beta)$
        \end{tabular}    \\ \hline
    S2 & \begin{tabular}{@{}c@{}} 
        $(a_5,\beta)$, \\
        $(a_3,\gamma)$, \\
        $(a_1,\gamma)$, \\ 
        $(a_4,\delta)$ 
        \end{tabular} \\ \hline
    \end{tabular}
    $\xrightarrow{\text{(a1)}} $
    \begin{tabular}{|c|}
        \hline
         ${R'}(A,B,D_R)$ \\ \hline
         \begin{tabular}{@{}c@{}} 
            $(a_1,\alpha,1)$, \\ 
            $(a_2,\alpha,2)$, \\
            $(a_3,\alpha,3)$, \\ 
            $(a_2,\beta,1)$
            \end{tabular}    \\ \hline
         \begin{tabular}{@{}c@{}} 
            $(a_5,\beta,2)$, \\
            $(a_3,\gamma,1)$, \\
            $(a_1,\gamma,2)$, \\ 
            $(a_4,\delta,1)$ 
            \end{tabular} \\ \hline
    \end{tabular}
    $\xrightarrow{\text{(a2)}} $
    \begin{tabular}{|c|}
        \hline
         ${R'}(A,B,D_R,D_S)$ \\ \hline
         \begin{tabular}{@{}c@{}} 
            $(a_1,\alpha,3,1)$, \\ 
            $(a_2,\alpha,3,1)$, \\
            $(a_3,\alpha,3,1)$, \\ 
            $(a_2,\beta,2,2)$
            \end{tabular}    \\ \hline
         \begin{tabular}{@{}c@{}} 
            $(a_5,\beta,2,2)$, \\
            $(a_3,\gamma,2,0)$, \\
            $(a_1,\gamma,2,0)$, \\ 
            $(a_4,\delta,1,3)$ 
            \end{tabular} \\ \hline
    \end{tabular}
 $\xrightarrow{\text{(b)}} $
    \begin{tabular}{|c|c|}
        \hline
        $\bar{R}(A,B)$  \\ \hline
        \begin{tabular}{@{}c@{}} 
            $(a_1,\alpha)$, \\ 
            $(a_2,\alpha)$, \\
            $(a_3,\alpha)$, \\ 
            $(a_2,\beta)$, \\
            $(a_2,\beta)$
        \end{tabular} \\ \hline
        \begin{tabular}{@{}c@{}} 
            $(a_5,\beta)$, \\
            $(a_5,\beta)$, \\
            $(a_4,\delta)$, \\
            $(a_4,\delta)$, \\
            $(a_4,\delta)$ 
        \end{tabular}  \\ \hline
    \end{tabular}
 $\xrightarrow{\text{(d)}} $
    \begin{tabular}{|c|c|c|}
        \hline
        ${T}(A,B,C)$  \\ \hline
        \begin{tabular}{@{}c@{}} 
            $(a_1,\alpha,c_1)$, \\ 
            $(a_2,\alpha,c_1)$, \\
            $(a_3,\alpha,c_1)$, \\ 
            $(a_2,\beta,c_1)$, \\
            $(a_2,\beta,c_2)$
        \end{tabular} \\ \hline
        \begin{tabular}{@{}c@{}} 
            $(a_5,\beta,c_1)$, \\
            $(a_5,\beta,c_2)$, \\
            $(a_4,\delta,c_3)$, \\
            $(a_4,\delta,c_4)$, \\
            $(a_4,\delta,c_5)$ 
        \end{tabular}  \\ \hline
    \end{tabular}
\hspace{-0.7em}
\multirow{2}{1em}{
\begin{tikzpicture}
\coordinate  (c0) at(-0.5,0) ;
\coordinate  (c1) at(0,0) ;
\coordinate  (c2) at(0,-5.0) ;
\coordinate  (c3) at(-0.38,-5.0) ;
\draw[<-] (c0) -- (c1);
\draw[-] (c1) -- (c2);
\draw[-] (c2) -- (c3) ;
\end{tikzpicture}
}
\bigskip\\

    \begin{tabular}{|c||c|}
    \hline
    & $S(B,C)$  \\ \hline
    S1 &  \begin{tabular}{@{}c@{}} 
                        $(\alpha,c_1)$, \\ 
                        $(\beta,c_1)$, \\
                        $(\beta,c_2)$, \\
                        $(\delta,c_3)$
                        \end{tabular}  \\ \hline
    S2 & \begin{tabular}{@{}c@{}} 
            $(\delta,c_4)$, \\ 
            $(\delta,c_5)$, \\
            $(\epsilon,c_1)$, \\
            $(\epsilon,c_2)$
            \end{tabular} \\ \hline
    \end{tabular}
    $\xrightarrow{\text{(a)}} $
    \begin{tabular}{|c|c|}
        \hline
          $S'(A,B,D_R,D_S)$ \\ \hline
         \begin{tabular}{@{}c@{}} 
                        $(\alpha,c_1,3,1)$, \\ 
                        $(\beta,c_1,2,2)$, \\
                        $(\beta,c_2,2,2)$, \\
                        $(\delta,c_3,1,3)$
                        \end{tabular}  \\ \hline
         \begin{tabular}{@{}c@{}} 
            $(\delta,c_4,1,3)$, \\ 
            $(\delta,c_5,1,3)$, \\
            $(\epsilon,c_1,0,2)$, \\
            $(\epsilon,c_2,0,2)$
            \end{tabular} \\ \hline
    \end{tabular}
 $\xrightarrow{\text{(b)}} $
    \begin{tabular}{|c|c|}
        \hline
        $\bar{S}(B,C,D_R,D_S,I,J,L,T)$  \\ \hline
        \begin{tabular}{@{}c@{}} 
            $(\alpha,c_1,3,1,1,1,1,1)$, \\ 
            $(\alpha,c_1,3,1,2,1,2,1)$, \\ 
            $(\alpha,c_1,3,1,3,1,3,1)$, \\ 
            $(\beta,c_1,2,2,1,4,4,1)$, \\
            $(\beta,c_1,2,2,2,4,6,2)$
        \end{tabular} \\ \hline
        \begin{tabular}{@{}c@{}} 
            $(\beta,c_2,2,2,3,4,5,1)$, \\
            $(\beta,c_2,2,2,4,4,7,2)$, \\
            $(\delta,c_3,1,3,1,8,8,2)$, \\
            $(\delta,c_4,1,3,2,8,9,2)$, \\ 
            $(\delta,c_5,1,3,3,8,10,2)$
        \end{tabular}  \\ \hline
    \end{tabular}
 $\xrightarrow{\text{(c)}} $
    \begin{tabular}{|c|c|}
        \hline
        $\bar{S}(B,C,\dots,L,T)$  \\ \hline
        \begin{tabular}{@{}c@{}} 
            $(\alpha,c_1,\dots,1,1)$, \\ 
            $(\alpha,c_1,\dots,2,1)$, \\ 
            $(\alpha,c_1,\dots,3,1)$, \\ 
            $(\beta,c_1,\dots,4,1)$, \\
            $(\beta,c_2,\dots,5,1)$
        \end{tabular} \\ \hline
        \begin{tabular}{@{}c@{}} 
            $(\beta,c_1,\dots,6,2)$, \\
            $(\beta,c_2,\dots,7,2)$, \\
            $(\delta,c_3,\dots,8,2)$, \\
            $(\delta,c_4,\dots,9,2)$, \\ 
            $(\delta,c_5,\dots,10,2)$
        \end{tabular}  \\ \hline
    \end{tabular}
\end{tabular}

\caption{Join algorithm example}
\label{fig:join}
\end{figure}

}
\end{example}
\section{Security Analysis}\label{sec:secure_analysis}

In this section, we prove that our proposed algorithms are both communication oblivious and computation oblivious. \rev{Recall Definition~\ref{def:comm} for} communication obliviousness. Note that all our algorithms involve communication only by calling the shuffle operator accompanied by determinate padding sizes, and the servers will receive messages whose sizes are congruent with these padding sizes $\{U_i\}$, which can be computed from the input sizes $\{n_i\}$ \rev{and public parameters $p$ and $\sigma$ (Theorem~\ref{thm:shuffle_by_key}, \ref{thm:expansion}, \ref{thm:join})}. Therefore, let the simulator simply outputs $\bar{S}$ as random numbers with sizes $\{U_i\}$, then the transcript of the algorithm and $\bar{S}$ \rev{are in identical sizes and hence indistinguishable}.

For computation obliviousness \rev{in Definition~\ref{def:comp}}, note that none of our algorithms involves any data dependent operations due to: (1) the execution of all loops with publicly known sizes; (2) the substitution of all conditional branches with $\cmove$ instructions; (3) the utilization of data independent memory access locations; (4) the employment of primitives that are intrinsically oblivious, \eg, $\osort$ and $\ocompact$. Therefore, the simulator can simply simulate the memory access pattern by running the algorithm with arbitrary input (of the same sizes), and the adversary could not distinguish between the access patterns from the true input and the simulated input. Note that for the random shuffle operator, the access pattern is random, but the distribution of the access pattern is data independent, thereby excluding any possibility for the adversary to differentiate.
\section{Evaluation}\label{sec:exp}

\subsection{Experimental Setup}
\paragraph{Environment} We deployed the distributed environment on 16 machines, each equipped with an Intel(R) Xeon(R) Platinum 8369B CPU @ 2.90GHz and having 1TB RAM capacity. For each machine, we initialized the enclave with 64GB \rev{EPC} size. 
The machines were interconnected via a local area network with bandwidth up to 2.9 GB/s, facilitating communication using the HTTP protocol built on Facebook's Proxygen framework~\cite{proxygen}. We counted the communication cost as the total number of bytes sent across the servers.
We chose AES-GCM with 128-bit key as the data encryption scheme, with encryption and decryption speed (inside enclave) around 1.0GB/s.
The compiler was GCC 8.5.0 with ``-O3'' optimization enabled, and the implementation of $\cmove$ followed the XOR-based C code in \cite{sp24}.
We enabled multi-threading outside the enclave, \ie, a server may receive data from other servers and perform computations inside the enclave simultaneously, but all computations within the enclave were executed using a single thread.

\paragraph{Default settings} For our experiments, we standardized the computational environment by configuring the number of servers to $p=16$ and setting the security parameter to $\sigma=40$. For join operator, we set $M$ to be the output size (no output padding). For a fair comparison, in evaluations involving shuffle or PK join operations, except when directly comparing these primitives, we consistently employed our implementations as described in Section~\ref{sec:basic}. This ensured that any observed performance differences could be attributed to the intrinsic merits of the algorithms under investigation rather than variations in the underlying primitives.

\subsection{Performance of Basic Operators}
\begin{figure}
    \includegraphics[width=\linewidth]{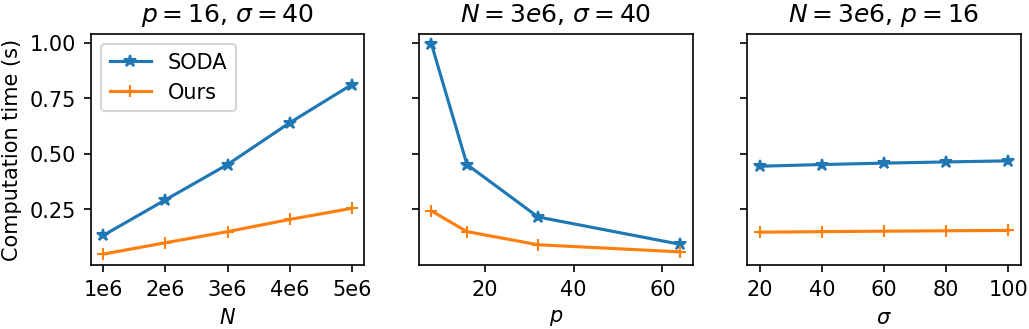}
    \caption{Computation time of $\opartition$ varying input size $N$, number of servers $p$, or security parameter $\sigma$.}
    \label{fig:exp_shuffle}
\end{figure}

\paragraph{OPartition} Our first improvement to the baseline is the standalone algorithm for $\opartition$, which is the key building block for the shuffle operator. We benchmarked the local computation phase of our algorithm against that of SODA~\cite{soda} using inputs generated randomly, varying input size $N$, the number of servers $p$, or the security parameter $\sigma$. The nature of obliviousness ensures that the performance of the algorithms remains consistent regardless of the variability in the input data.  The results are shown in Figure~\ref{fig:exp_shuffle}. 
% To indicate that the computation time is nearly inversely proportional to $p$, we use $p^{-1}$ for the x-axis.  
Our algorithm shows a substantial empirical performance improvement, ranging from 60\% to as much as 310\%. Notably, this enhancement factor grows in proportion to the increase in input size $N$ or the decrease in the number of servers $p$. This trend is in line with our theoretical analysis, which states an improvement factor on the order of $\log n/\log p$.  Besides, we find that both algorithms are insensitive to the security parameter. In applications requiring even smaller failure probability, the performance degradation will be minimal.

\paragraph{Primary key join (PK join)} We conducted the PK join experiments on the well-known TPC-H dataset evaluating our Algorithm~\ref{alg:pkjoin} against Opaque's PK join and SODA's join using the query below:
\lstset{breaklines=false}
\begin{lstlisting}[mathescape=true]
SELECT * FROM orders JOIN customer ON o_custkey$\,=\,$c_custkey;
\end{lstlisting}
Note that \texttt{c\_custkey} is the primary key of table \texttt{customer}, so the query is a PK join. Both tables were generated by the code from a publicly accessible GitHub repository \cite{TPCHSKEW} with a scale factor of 10 and with the \texttt{orders} table exhibiting four distinct values of \rev{Zipfian distribution parameters}, denoted by $z$, as outlined in Table~\ref{tab:exp_pkjoin_input}. Irrelevant columns were eliminated from computation, leaving only \texttt{c\_custkey}, \texttt{c\_nationkey}, \texttt{o\_orderkey}, and \texttt{o\_custkey}. 

\begin{table}
    \caption{PK join input table information; Total time and communication cost of the PK join algorithms}\label{tab:exp_pkjoin_input}
    \centering
    \begin{tabular}{lrrrrrr} \toprule
     & \multicolumn{2}{c}{\texttt{customer}} & \multicolumn{4}{c}{\texttt{orders}}  \\ \cmidrule(lr){2-3}\cmidrule(lr){4-7}
    \#Rows   & \multicolumn{2}{c}{$N_2=1.5 \!\times\! 10^6$} &  \multicolumn{4}{c}{$N_1=1.5\!\times\! 10^7$}\\ 
    % Input size   & \multicolumn{2}{|c|}{0.017GB} &  \multicolumn{4}{|c|}{0.17GB}\\ \hline
    \rev{Zipf parameter} $z$ & \multicolumn{2}{c}{0} & 0 & 0.5 & 1 & 1.5 \\%\hline 
    Max degree $\alpha$ & \multicolumn{2}{c}{1} & 37 & 449 & 50697 & 210490  \\ \midrule %\hline
    & Ours & Opaque & \multicolumn{4}{c}{SODA} \\ \cmidrule(lr){2-2}\cmidrule(lr){3-3}\cmidrule(lr){4-7}
    \rev{\#Output ($\times 10^7$)} & \rev{1.50} & \rev{1.50} & \rev{1.50} & \rev{1.50} & \rev{1.58} & \rev{1.84} \\ %\hline
    Total time (s) & 6.44 & 12.5 & 11.0 & 11.0 & 11.3 & 12.0 \\ %\hline
    Comm. cost (GB) & 0.62 & 1.56 & 1.19 & 1.19 & 1.23 & 1.34 \\ \bottomrule
    \end{tabular}
\end{table}

The results are also in Table~\ref{tab:exp_pkjoin_input}. Different \rev{values of $z$} result in different maximum degrees on the join key, thereby affecting the \rev{output size} of SODA's join but not impacting our algorithm or Opaque's join. 
In terms of overall running time, our algorithm improves upon the baselines by at least 70\%. This improvement rises to more than 90\% regarding communication costs. While SODA's join slightly outperforms Opaque's PK join, the gap narrows as \rev{$z$} increases, \rev{due to the increasing output size}. Meanwhile, both Opaque's PK join and our algorithm maintain steady performance despite varying \rev{$z$} due to obliviousness.  

\subsection{Performance of Join}  
For join, in addition to evaluating \algname and SODA, we also conducted tests on a single server using the state-of-the-art oblivious standalone join~\cite{10.14778/3407790.3407814}, in which all data is sent to the first server that performs the standalone join locally and then sends the results to the other servers.

\paragraph{Varying datasets}
We evaluated the join operator on Stanford Large Network Dataset Collection~\cite{snapnets}. We selected four graphs with various \rev{$\ell_\infty$-skewnesses}: ``com-DBLP'' (\textbf{DBLP}), ``email-EuAll'' (\textbf{email}), ``com-Youtube'' (\textbf{Youtube}), and ``wiki-topcats'' (\textbf{wiki}). More information of the four graphs are in Table~\ref{tab:exp_join_input}. Each graph was converted into a relational table format with two columns, \texttt{src} and \texttt{dst}, to represent the source and destination nodes of each edge. 
We assessed the performance on the following self-join query designed to identify all length-2 paths within the graphs:
\begin{lstlisting}[mathescape=true]
    SELECT * FROM graph R JOIN graph S ON R.dst$\,=\,$S.src;
\end{lstlisting}

\begin{table}
    \caption{Join input table information; Output sizes of the join algorithms}\label{tab:exp_join_input}
    \centering
    \begin{tabular}{lrrrr} \toprule
    & \textbf{DBLP} & \textbf{email}  & \textbf{Youtube} & \textbf{wiki} \\\midrule
    \#Input $N_1=N_2$ & $1.0\!\times\! 10^6$ & $4.2\!\times\! 10^5$ &  $2.9\!\times\! 10^6$ & $2.9\!\times\! 10^7$ \\ % \hline
    % Total I/O size  & 0.13GB & 0.75GB  & 2.90GB & 39.4GB \\\hline
    Max \#dst $\alpha_1$ & 306 & 930  &   28576 & 3907\\ %\hline
    Max \#src $\alpha_2$ & 113 & 7631  &   4256 & 238040\\ %\hline
    \rev{$\ell_\infty$-skewness $\phi$} &  \rev{$0.0049$} &  \rev{0.14} & \rev{0.64} & \rev{0.36} \\ %\hline
    \midrule
    \#Output $M$  & $7.1\!\times\! 10^6$ & $5.0\!\times\! 10^7$  & $1.9\!\times\! 10^8$ & $2.6\!\times\! 10^9 $ \\%\hline
    \rev{\#Output (SODA)} &  \rev{$7.7\!\times\! 10^6$} & \rev{$16\!\times\! 10^7$} & \rev{$21\!\times\! 10^8$}  & \rev{$17\!\times\! 10^9$} \\ \bottomrule
    \end{tabular}
\end{table}

\begin{figure*}
    \centering
    \begin{subfigure}[t]{0.32\textwidth}
        \centering
        \includegraphics[width=\linewidth]{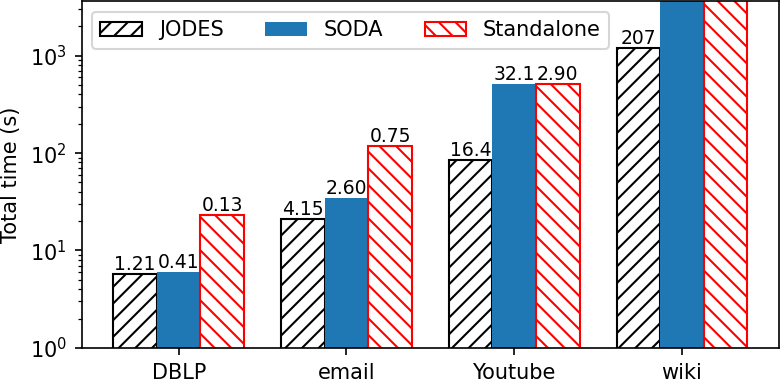}
        \caption{Varying input datasets}
        \label{fig:exp_join}
    \end{subfigure}
    \hfill
    \begin{subfigure}[t]{0.213\textwidth}
        \centering
        \includegraphics[width=\linewidth]{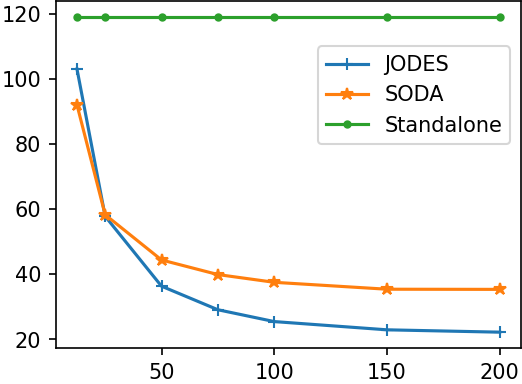}
        \caption{Varying bandwidths (MB/s)}
        \label{fig:exp_limit_band_join}
    \end{subfigure}
    \hfill
    \begin{subfigure}[t]{0.213\textwidth}
        \centering
        \includegraphics[width=\linewidth]{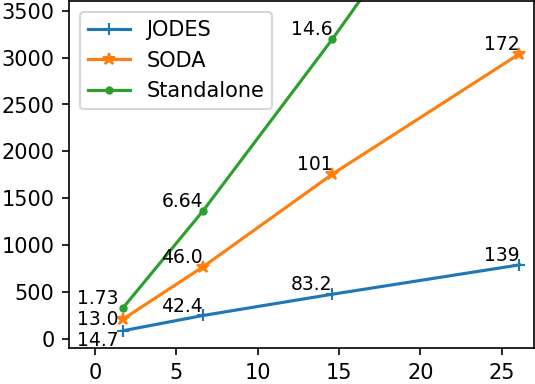}
        \caption{Varying I/O sizes (GB)}
        \label{fig:exp_sample_join}
    \end{subfigure}
    \hfill
    \begin{subfigure}[t]{0.213\textwidth}
        \centering
        \includegraphics[width=\linewidth]{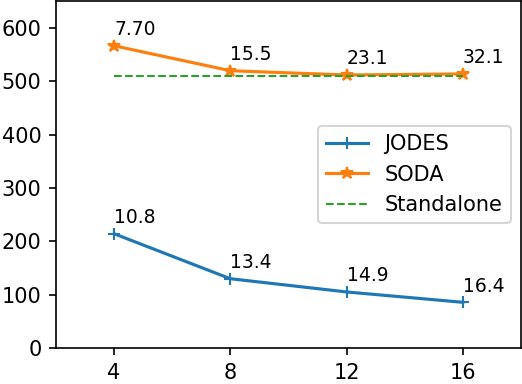}
        \caption{Varying number of servers}
        \label{fig:exp_nodes_join}
    \end{subfigure}
    \caption{Total time of join, where the labels on top of the bars or adjacent to the data points are communication costs (GB).}
    \label{fig:combined_figures}
\end{figure*}

Figure~\ref{fig:exp_join} includes the performance results, with the y-axis represented on a logarithmic scale. For the \textbf{wiki} dataset, both SODA and standalone algorithm could not complete within an hour. The speed-up of \algname compared to \rev{the standalone algorithm} ranges from 4x (for small data) to 6x (for large data). In contrast, our speed-up over SODA is highly dependent on the value of \rev{$\phi$}: it is 1.1x for \textbf{DBLP} (small \rev{$\phi$}), 1.6x for \textbf{email} (medium \rev{$\phi$}), and 6x for \textbf{Youtube} (large \rev{$\phi$}). Specifically, for the \textbf{Youtube} dataset, the total time of SODA using 16 servers even exceeds that of \rev{the standalone algorithm} with only one server, thus losing the advantages of distribution. 
With respect to communication costs, \rev{the standalone algorithm} incurs the least, equivalent to only the I/O size. \algname exhibits slightly higher communication costs than SODA for \textbf{DBLP} and \textbf{email}, but lower costs for \textbf{Youtube}.

% \begin{figure}
%     \includegraphics[width=\linewidth]{figures/join.png}
%     \caption{Total time of join varying input datasets, where the labels on top of the bars are communication costs (GB).}
%     \label{fig:exp_join}
% \end{figure}

\paragraph{Varying bandwidths}
We note that for \textbf{DBLP} and \textbf{email} in Figure~\ref{fig:exp_join}, \algname has a shorter running time but incurs a higher communication costs than SODA. \rev{The reason is that SODA adopts complex design of local computations to minimize the communication cost. To see whether SODA outperforms \algname for limited bandwidth,} we reran the tests on the \textbf{email} dataset and limited the bandwidth to assess its impact on performance. The results are shown in Figure~\ref{fig:exp_limit_band_join}. \rev{Since varying bandwidths does not change the communication costs, we only plot the running time.}  It appears that %\algname's performance is more sensitive to bandwidth. 
only under very limited bandwidth conditions (less than 25 MB/s), \algname's performance is surpassed by SODA. However, we argue that in distributed settings, a \rev{large bandwidth requirement is reasonable}. For example, all instances of Amazon EMR~\cite{AWSEMR} have a minimum bandwidth of 10 Gbps (\rev{1280 MB/s}).

% \paragraph{Varying bandwidths}
% We note that for \textbf{DBLP} and \textbf{email} in Figure~\ref{fig:exp_join}, \algname has a shorter running time but incurs a higher communication costs than SODA. \rev{The reason is that SODA adopts complex design of location computations to minimize the communication cost. To see whether SODA outperforms \algname for limited bandwidth,} we reran the tests on the \textbf{email} dataset and limited the bandwidth to assess its impact on performance. The results are shown in Figure~\ref{fig:exp_limit_band_join}. \rev{Since varying bandwidths does not change the communication costs, we only plots the running time.}  It appears that %\algname's performance is more sensitive to bandwidth. 
% only under very limited bandwidth conditions (less than 25 MB/s), \algname's performance is surpassed by SODA. However, we argue that in distributed settings, a \rev{large bandwidth requirement is reasonable}. For example, all instances of Amazon EMR~\cite{AWSEMR} have a minimum bandwidth of 10 Gbps (\rev{1280 MB/s}).

% \begin{figure}
%     \includegraphics[width=\linewidth]{figures/limit_band_join.png}
%     \caption{Total time of join varying bandwidth.}
%     \label{fig:exp_limit_band_join}
% \end{figure}

\paragraph{Varying I/O sizes}
We also conducted experiments on sampled data from the \textbf{wiki} dataset to assess the scalability of \algname and SODA. Specifically, we sampled each row of the input table with probability $\epsilon$ independently and then performed the join on the sampled table, as described by the following SQL:
\begin{lstlisting}[mathescape=true]
WITH sampled AS (SELECT * FROM graph WHERE rand() < $\epsilon$)
SELECT * FROM sampled R JOIN sampled S ON R.dst$\,=\,$S.src;
\end{lstlisting}
Note that a sampling probability of $\epsilon$ induces the expected join size of the sampled table to be $\epsilon^2$ times that of the original table. We tried $\epsilon\in\{0.2,0.4,0.6,0.8\}$, and the performance results are shown in Figure~\ref{fig:exp_sample_join}. For $\epsilon=0.8$, \rev{the standalone algorithm} could not finish in an hour.  The total time of all algorithms scales almost linearly to the \rev{I/O sizes, \ie, the total sizes of input and output}. The speed-up factor of \algname to SODA ranges from 2.5 ($\epsilon=0.2)$ to 3.9 ($\epsilon=0.8$). \algname also incurs less communication cost, except when $\epsilon=0.2$, where it is slightly higher.

\rev{
\paragraph{Varying number of servers}
To examine how the number of servers influences performance, we conducted experiments with different values of $p$ using the \textbf{Youtube} dataset, and the results are presented in Figure~\ref{fig:exp_nodes_join}. We observe that SODA does not scale well for large $p$: the running time when using 12 servers is almost identical to that when using 16 servers. The major reason is that the output size of SODA on each server is $M(1/p+\phi)$. Since the extra padding size $M\phi$ is independent of $p$ and dominates the join size $M/p$, increasing the number of servers does not significantly reduce the workload on each server, while the increasing total communication can even degrade the performance. On the contrary, the output size of \algname on each server is always $M/p$, allowing it to scale much more effectively.
}

\rev{
\paragraph{Running time breakdown analysis}
\algname (Algorithm~\ref{alg:join}) works in a manner similar to the standalone algorithm at a high level, which can be segmented into the following three phases:
\begin{enumerate}
    \item \textit{Preparation.} This phase involves computing the degrees of join keys in both input tables and matching them through a primary key (PK) join (Lines~\ref{line:preparation_start}--\ref{line:preparation_end});
    \item \textit{Expansion.} In this phase, both tables are expanded from size $N$ to $M$ through expansion operations (Lines~\ref{line:expansion_start}--\ref{line:expansion_end});
    \item \textit{Alignment.} This phase focuses on aligning the two tables to ensure that the correct tuples match (Lines~\ref{line:computeL_start}--\ref{line:alignment_end}).
\end{enumerate}
% We tested the breakdown running time on the three parts over the \textbf{Youtube} dataset, and the results are in Table~\ref{tab:breakdown}. We observe that the preparation time is minor, as it only involves in operations on input tables with $O(N)$ size. In contrast, both expansion and alignment generate or operate on the output table with $O(M)$ size, with $M\gg N$. Using 16 servers, the improvement factors of \algname to the standalone algorithm on the three parts are 5.2, 3.5, 7.7, respectively. This result indicates that, to further improve \algname, a good direction could be optimizing the design of the expansion algorithm.

We evaluated the breakdown of running time for these three phases using the \textbf{YouTube} dataset, and the results are presented in Table~\ref{tab:breakdown}. The findings align with our theoretical analysis, indicating that the preparation phase depends solely on $N$, while both the expansion and alignment phases are influenced by $M$, where $M \gg N$. Utilizing 16 servers, the performance improvement factors of \algname compared to the standalone algorithm for the three phases are 5.2, 3.5, and 7.7, respectively. This suggests that optimizing the design of the expansion algorithm could be a productive pathway for further enhancing \algname.
}

% Youtube dataset
% Jodes 85.572s (100%)
% Single 511s (597%)
% prepare1:6620.92 ms
% prepare2:9075.57 ms
% expand:54138.2 ms
% prepare2:9023.14 ms
% expand:54423.1 ms
% ---
% prepare: 24.72s
% expand: 108.56s
% sorting: 366.42s
% Total: 499.7s
% ---
% factor: 511/499.7/85.572=
% prepare: 29.5%
% expand: 129.7%
% sorting: 437.9%
% total: 597.1%

\begin{table}
    \caption{\rev{Running time breakdown of \algname and the standalone algorithm}}\label{tab:breakdown}
    \centering
    \begin{tabular}{ccccc} 
    \toprule
    % \textbf{Algorithm} 
    & \textbf{Preparation} & \textbf{Expansion} & \textbf{Alignment} & \textbf{Total} \\
    \midrule
    \algname      & 5.7\% & 37.2\% & 57.1\% & 100\% \\
    Standalone    & 29.5\% & 129.7\% & 437.9\% & 597\% \\
    \bottomrule
    \end{tabular}
\end{table}
\section{Related Work}
Existing analytic systems based on TEE include \cite{10.1145/3318464.3386141,10.14778/3364324.3364331,opaque,Gribov2017StealthDBAS,10.14778/3447689.3447705,10.14778/3554821.3554826,10.14778/3583140.3583158}. However, most of them only focus on the standalone setting.
\citet{10.1145/2810103.2813695} firstly pointed out the leakage by the network traffic in the distributed setting. \rev{Their empirical analysis on datasets that include personal and geographical data shows that the runs of typical jobs can infer precise information about their input. To prevent such leakage,}  they provided a shuffle-in-the middle solution: before sending data (with some padding techniques) to the intended destination, permuting the input randomly among the servers in advance to remove potential skewness; \citet{mpc4mpc} proposed a different solution based on oblivious routing, \rev{which packs data into several bins, and routes the bins to a random server through the butterfly network}. Both solutions turn any non-oblivious algorithm to the oblivious counterpart. Nevertheless, the communication cost blows up by a constant factor (at least 2) only when the load of the non-oblivious algorithm is balanced, \ie, the number of elements any server received in any round is $O(R/p)$ where $R$ is the total number of elements received of all the $p$ servers. Without load balancing,  the communication cost can increase by a factor of up to $p$ in the worst-case scenario. However, existing non-oblivious join algorithms, including the hash join and sort merge join adopted by Spark~\cite{10.5555/2228298.2228301}, do not satisfy the constraint. Note that even in the plaintext model where obliviousness is not required, such imbalance happens when the input is skewed, leading to severe performance downgrade.
Our join algorithm \algname naturally provides a solution to this issue caused by input skew in the plaintext model, because its performance is independent of the input and hence its skewness.

Opaque~\cite{opaque} proposes an encrypted distributed analytic system based on Spark. Unlike these general solutions, it designs specialized oblivious algorithms for sorting, filter, aggregate, and PK join, but not (general equi-)join. Most of their designs are based on its oblivious sorting, which is implemented based on column sort. SODA~\cite{soda} considers column sort to be heavy, so it proposes its own oblivious algorithms for filter, aggregate, and join without relying on oblivious sorting, but SODA's join needs to publicize the maximum degrees of the input tables.

A circuit also naturally induces an oblivious algorithm in the distributed setting. To evaluate the circuit, it is necessary to ensure the inputs of each gate lie on the same server. Therefore, it incurs a communication round before each level, hence the number of rounds of the algorithm is linear to the depth of the circuit. However, existing circuits for database joins are all with $\Omega(\log^2 N)$ depth~\cite{10.1145/3517804.3524142,10.14778/3407790.3407814} and hence will induce algorithms with polylogarithm number of communication rounds, which severely downgrades the performance due to network latency. Meanwhile, \rev{common distributed} algorithms introduced \rev{above} incur only $O(1)$ communication rounds.

%\iffull
\rev{
\paragraph{Plaintext distributed join} Numerous studies have been conducted on join algorithms within the plaintext distributed model, as referenced in~\cite{10.1145/3125644,10.1145/3034786.3034788,10.1145/3294052.3319698,10.1145/3375395.3387657,10.1145/3452021.3458319}. These studies primarily focus on the massively parallel computation model where only the sizes of data received are considered, while disregarding the costs of sending data (including emitting the output) and local computations. However, these algorithms are unsuitable for cloud-based encrypted systems because their local computations, when made oblivious, incur costs that are significantly higher than negligible. Furthermore, the sizes of local joins in their final rounds are data-dependent, which presents challenges for adapting them into oblivious algorithms.}

% However, these algorithms are not suitable for a cloud-based encrypted system, as their local computation, when made oblivious, should have cost much larger than negligible. Moreover, the local join sizes at the final round of them are also data dependent, which poses an challenge for turning them into oblivious algorithms.

% To maximize the benefits of this model, the algorithm in~\cite{10.1145/3034786.3056110} postpones the most heavy execution of local joins, which increase the input size to $M$, until the final round, and then directly emit the output with no cost. However, MPC model is not suitable for a cloud-based encrypted system, as local computation, particularly when made oblivious, should have cost much larger than negligible. Moreover, the local join sizes at the final round  are also data dependent. Making this sizes data dependent would require oversized padding and make the algorithm impractical.
%\iffull
\paragraph{Join under MPC} There are several join algorithms \cite{fastdatabasejoin,10.14778/3055330.3055334,secrecy,secretsharedjoins,DBLP:conf/icde/HanZFLL22,10.1145/3448016.3452808,263796} under the \textit{secure multi-party computation} (MPC) model, in which several servers jointly compute the join over the secret shared data from the user. The security guarantee of MPC is incomparable to the distributed TEE model: Under MPC, the user does not need to trust any hardware as in TEE, but they believe that the servers will not collude to steal data from the user. In  real-world scenarios, the servers in distributed TEE can belong to the same cluster connected by network with low latency and high bandwidth, while servers in MPC are usually from different organizations (\eg, Alibaba, Amazon, and Azure). Regarding efficiency, the speed of join under MPC is usually slower than the one in the standalone TEE setting, which is slower than the distributed TEE setting,  All existing join algorithms under MPC incur both computation and communication cost $\Omega(N\log N+M)$ with a considerable hidden constant factor.

%Several join algorithms exist within the framework of secure multi-party computation (MPC) \cite{fastdatabasejoin,10.14778/3055330.3055334,secrecy,secretsharedjoins,DBLP:conf/icde/HanZFLL22,10.1145/3448016.3452808,263796}, where multiple servers hold the secret shares from the user and collaboratively execute a predetermined MPC protocol. The security offered by MPC is distinct from that in the distributed TEE model. In an MPC setting, users are not required to place their trust in any particular hardware as they would with TEEs. Instead, they must trust that the servers will not collude to steal the user's data. In real-world scenarios, servers operating within a distributed TEE may belong to a unified cluster benefiting from a network characterized by low latency and high bandwidth. In contrast, MPC servers often represent disparate organizations (\eg, Alibaba, Amazon, and Azure). In terms of efficiency, join algorithms under MPC tend to be slower than those in a standalone TEE setting, and a standalone TEE, in turn, is expected to be slower than a distributed TEE configuration. In theory, all existing join algorithms under MPC incur computational and communication costs of $\Omega(N\log N+M)$, each accompanied by a significant hidden constant factor.
%\fi
% \else
% Related work on plaintext distributed joins and joins under MPC is discussed in the full version of this paper \cite{full}.
% \fi

\section{Conclusion and Future Work}
We have proposed \algname, an oblivious algorithm in the distributed setting that is superior to existing works in both theoretical and experimental aspects. 
Following the idea in~\cite{10.1145/3034786.3056110}, one can prove that the communication cost of a perfect load balanced oblivious join (\ie, each server holds $O(M/p)$ of the output tuples of join result) is $\Omega(N+\sqrt{Mp})$. Since the communication costs of existing oblivious join algorithms are all $\Omega(N+M)$, an interesting future research direction is to close the gap, \ie, either proposing an oblivious join algorithm with less cost or providing a stronger lower bound. 

\bibliographystyle{plainnat}
\bibliography{main}

\begin{thebibliography}{62}
\providecommand{\natexlab}[1]{#1}
\providecommand{\url}[1]{\texttt{#1}}
\expandafter\ifx\csname urlstyle\endcsname\relax
  \providecommand{\doi}[1]{doi: #1}\else
  \providecommand{\doi}{doi: \begingroup \urlstyle{rm}\Url}\fi

\bibitem[Afrati and Ullman(2011)]{5710932}
Foto~N. Afrati and Jeffrey~D. Ullman.
\newblock Optimizing multiway joins in a map-reduce environment.
\newblock \emph{IEEE Transactions on Knowledge and Data Engineering}, 23\penalty0 (9):\penalty0 1282--1298, 2011.
\newblock \doi{10.1109/TKDE.2011.47}.

\bibitem[Aggarwal and Vitter(1988)]{10.1145/48529.48535}
Alok Aggarwal and S.~Vitter, Jeffrey.
\newblock The input/output complexity of sorting and related problems.
\newblock \emph{Commun. ACM}, 31\penalty0 (9):\penalty0 1116–1127, sep 1988.
\newblock ISSN 0001-0782.
\newblock \doi{10.1145/48529.48535}.
\newblock URL \url{https://doi.org/10.1145/48529.48535}.

\bibitem[Ajtai et~al.(1983)Ajtai, Koml\'{o}s, and Szemer\'{e}di]{10.1145/800061.808726}
M.~Ajtai, J.~Koml\'{o}s, and E.~Szemer\'{e}di.
\newblock An 0(n log n) sorting network.
\newblock In \emph{Proceedings of the Fifteenth Annual ACM Symposium on Theory of Computing}, STOC '83, page 1–9, New York, NY, USA, 1983. Association for Computing Machinery.
\newblock ISBN 0897910990.
\newblock \doi{10.1145/800061.808726}.
\newblock URL \url{https://doi.org/10.1145/800061.808726}.

\bibitem[Antonopoulos et~al.(2020)Antonopoulos, Arasu, Singh, Eguro, Gupta, Jain, Kaushik, Kodavalla, Kossmann, Ogg, Ramamurthy, Szymaszek, Trimmer, Vaswani, Venkatesan, and Zwilling]{10.1145/3318464.3386141}
Panagiotis Antonopoulos, Arvind Arasu, Kunal~D. Singh, Ken Eguro, Nitish Gupta, Rajat Jain, Raghav Kaushik, Hanuma Kodavalla, Donald Kossmann, Nikolas Ogg, Ravi Ramamurthy, Jakub Szymaszek, Jeffrey Trimmer, Kapil Vaswani, Ramarathnam Venkatesan, and Mike Zwilling.
\newblock Azure sql database always encrypted.
\newblock In \emph{Proceedings of the 2020 ACM SIGMOD International Conference on Management of Data}, SIGMOD '20, page 1511–1525, New York, NY, USA, 2020. Association for Computing Machinery.
\newblock ISBN 9781450367356.
\newblock \doi{10.1145/3318464.3386141}.
\newblock URL \url{https://doi.org/10.1145/3318464.3386141}.

\bibitem[Arasu and Kaushik(2014)]{DBLP:conf/icdt/ArasuK14}
Arvind Arasu and Raghav Kaushik.
\newblock Oblivious query processing.
\newblock In Nicole Schweikardt, Vassilis Christophides, and Vincent Leroy, editors, \emph{Proc. 17th International Conference on Database Theory (ICDT), Athens, Greece, March 24-28, 2014}, pages 26--37. OpenProceedings.org, 2014.
\newblock \doi{10.5441/002/ICDT.2014.07}.
\newblock URL \url{https://doi.org/10.5441/002/icdt.2014.07}.

\bibitem[Arasu et~al.(2013)Arasu, Blanas, Eguro, Joglekar, Kaushik, Kossmann, Ramamurthy, Upadhyaya, and Venkatesan]{10.1145/2463676.2467797}
Arvind Arasu, Spyros Blanas, Ken Eguro, Manas Joglekar, Raghav Kaushik, Donald Kossmann, Ravi Ramamurthy, Prasang Upadhyaya, and Ramarathnam Venkatesan.
\newblock Secure database-as-a-service with cipherbase.
\newblock In \emph{Proceedings of the 2013 ACM SIGMOD International Conference on Management of Data}, SIGMOD '13, page 1033–1036, New York, NY, USA, 2013. Association for Computing Machinery.
\newblock ISBN 9781450320375.
\newblock \doi{10.1145/2463676.2467797}.
\newblock URL \url{https://doi.org/10.1145/2463676.2467797}.

\bibitem[Asharov et~al.(2020)Asharov, Chan, Nayak, Pass, Ren, and Shi]{doi:10.1137/1.9781611976014.2}
Gilad Asharov, TH~Hubert Chan, Kartik Nayak, Rafael Pass, Ling Ren, and Elaine Shi.
\newblock Bucket oblivious sort: An extremely simple oblivious sort.
\newblock In \emph{Symposium on Simplicity in Algorithms}, pages 8--14. SIAM, 2020.
\newblock \doi{10.1137/1.9781611976014.2}.
\newblock URL \url{https://epubs.siam.org/doi/abs/10.1137/1.9781611976014.2}.

\bibitem[Badrinarayanan et~al.(2022)Badrinarayanan, Das, Garimella, Raghuraman, and Rindal]{secretsharedjoins}
Saikrishna Badrinarayanan, Sourav Das, Gayathri Garimella, Srinivasan Raghuraman, and Peter Rindal.
\newblock Secret-shared joins with multiplicity from aggregation trees.
\newblock In \emph{Proceedings of the 2022 ACM SIGSAC Conference on Computer and Communications Security}, page 209–222. Association for Computing Machinery, 2022.

\bibitem[Batcher(1968)]{10.1145/1468075.1468121}
K.~E. Batcher.
\newblock Sorting networks and their applications.
\newblock In \emph{Proceedings of the April 30--May 2, 1968, Spring Joint Computer Conference}, AFIPS '68 (Spring), page 307–314, New York, NY, USA, 1968. Association for Computing Machinery.
\newblock ISBN 9781450378970.
\newblock \doi{10.1145/1468075.1468121}.
\newblock URL \url{https://doi.org/10.1145/1468075.1468121}.

\bibitem[Bater et~al.(2017)Bater, Elliott, Eggen, Goel, Kho, and Rogers]{10.14778/3055330.3055334}
Johes Bater, Gregory Elliott, Craig Eggen, Satyender Goel, Abel Kho, and Jennie Rogers.
\newblock Smcql: Secure querying for federated databases.
\newblock \emph{Proc. VLDB Endow.}, 10\penalty0 (6):\penalty0 673–684, feb 2017.
\newblock ISSN 2150-8097.
\newblock \doi{10.14778/3055330.3055334}.
\newblock URL \url{https://doi.org/10.14778/3055330.3055334}.

\bibitem[Bater et~al.(2018)Bater, He, Ehrich, Machanavajjhala, and Rogers]{10.14778/3291264.3291274}
Johes Bater, Xi~He, William Ehrich, Ashwin Machanavajjhala, and Jennie Rogers.
\newblock Shrinkwrap: Efficient sql query processing in differentially private data federations.
\newblock \emph{Proc. VLDB Endow.}, 12\penalty0 (3):\penalty0 307–320, nov 2018.
\newblock ISSN 2150-8097.
\newblock \doi{10.14778/3291264.3291274}.
\newblock URL \url{https://doi.org/10.14778/3291264.3291274}.

\bibitem[Beame et~al.(2014)Beame, Koutris, and Suciu]{10.1145/2594538.2594558}
Paul Beame, Paraschos Koutris, and Dan Suciu.
\newblock Skew in parallel query processing.
\newblock In \emph{Proceedings of the 33rd ACM SIGMOD-SIGACT-SIGART Symposium on Principles of Database Systems}, PODS '14, page 212–223, New York, NY, USA, 2014. Association for Computing Machinery.
\newblock ISBN 9781450323758.
\newblock \doi{10.1145/2594538.2594558}.
\newblock URL \url{https://doi.org/10.1145/2594538.2594558}.

\bibitem[Beame et~al.(2017)Beame, Koutris, and Suciu]{10.1145/3125644}
Paul Beame, Paraschos Koutris, and Dan Suciu.
\newblock Communication steps for parallel query processing.
\newblock \emph{J. ACM}, 64\penalty0 (6), oct 2017.
\newblock ISSN 0004-5411.
\newblock \doi{10.1145/3125644}.
\newblock URL \url{https://doi.org/10.1145/3125644}.

\bibitem[Chan et~al.(2020)Chan, Chung, Lin, and Shi]{mpc4mpc}
T-H.~Hubert Chan, Kai-Min Chung, Wei-Kai Lin, and Elaine Shi.
\newblock {MPC for MPC: Secure Computation on a Massively Parallel Computing Architecture}.
\newblock In Thomas Vidick, editor, \emph{11th Innovations in Theoretical Computer Science Conference (ITCS 2020)}, volume 151 of \emph{Leibniz International Proceedings in Informatics (LIPIcs)}, pages 75:1--75:52, Dagstuhl, Germany, 2020. Schloss Dagstuhl--Leibniz-Zentrum fuer Informatik.
\newblock ISBN 978-3-95977-134-4.
\newblock \doi{10.4230/LIPIcs.ITCS.2020.75}.
\newblock URL \url{https://drops.dagstuhl.de/opus/volltexte/2020/11760}.

\bibitem[Chang et~al.(2022)Chang, Xie, Wang, and Li]{10.1145/3514221.3517868}
Zhao Chang, Dong Xie, Sheng Wang, and Feifei Li.
\newblock Towards practical oblivious join.
\newblock In \emph{Proceedings of the 2022 International Conference on Management of Data}, SIGMOD '22, page 803–817, New York, NY, USA, 2022. Association for Computing Machinery.
\newblock ISBN 9781450392495.
\newblock \doi{10.1145/3514221.3517868}.
\newblock URL \url{https://doi.org/10.1145/3514221.3517868}.

\bibitem[Dean and Ghemawat(2008)]{10.1145/1327452.1327492}
Jeffrey Dean and Sanjay Ghemawat.
\newblock Mapreduce: Simplified data processing on large clusters.
\newblock \emph{Commun. ACM}, 51\penalty0 (1):\penalty0 107–113, jan 2008.
\newblock ISSN 0001-0782.
\newblock \doi{10.1145/1327452.1327492}.
\newblock URL \url{https://doi.org/10.1145/1327452.1327492}.

\bibitem[Dwork and Roth(2014)]{10.1561/0400000042}
Cynthia Dwork and Aaron Roth.
\newblock The algorithmic foundations of differential privacy.
\newblock \emph{Found. Trends Theor. Comput. Sci.}, 9\penalty0 (3–4):\penalty0 211–407, aug 2014.
\newblock ISSN 1551-305X.
\newblock \doi{10.1561/0400000042}.
\newblock URL \url{https://doi.org/10.1561/0400000042}.

\bibitem[El-Hindi et~al.(2022)El-Hindi, Ziegler, Heinrich, Lutsch, Zhao, and Binnig]{10.1145/3533737.3535098}
Muhammad El-Hindi, Tobias Ziegler, Matthias Heinrich, Adrian Lutsch, Zheguang Zhao, and Carsten Binnig.
\newblock Benchmarking the second generation of intel sgx hardware.
\newblock DaMoN '22, New York, NY, USA, 2022. Association for Computing Machinery.
\newblock ISBN 9781450393782.
\newblock \doi{10.1145/3533737.3535098}.
\newblock URL \url{https://doi.org/10.1145/3533737.3535098}.

\bibitem[Eskandarian and Zaharia(2019)]{10.14778/3364324.3364331}
Saba Eskandarian and Matei Zaharia.
\newblock Oblidb: Oblivious query processing for secure databases.
\newblock \emph{Proc. VLDB Endow.}, 13\penalty0 (2):\penalty0 169–183, oct 2019.
\newblock ISSN 2150-8097.
\newblock \doi{10.14778/3364324.3364331}.
\newblock URL \url{https://doi.org/10.14778/3364324.3364331}.

\bibitem[Evans et~al.(2018)Evans, Kolesnikov, and Rosulek]{8584398}
David Evans, Vladimir Kolesnikov, and Mike Rosulek.
\newblock A pragmatic introduction to secure multi-party computation.
\newblock \emph{Found. Trends Priv. Secur.}, 2\penalty0 (2–3):\penalty0 70–246, dec 2018.
\newblock ISSN 2474-1558.
\newblock \doi{10.1561/3300000019}.
\newblock URL \url{https://doi.org/10.1561/3300000019}.

\bibitem[Facebook(2022)]{proxygen}
Facebook.
\newblock Proxygen: Facebook's c++ http libraries, 2022.
\newblock URL \url{https://github.com/facebook/proxygen/releases/tag/v2022.11.14.00}.

\bibitem[Goldreich and Ostrovsky(1996)]{10.1145/233551.233553}
Oded Goldreich and Rafail Ostrovsky.
\newblock Software protection and simulation on oblivious rams.
\newblock \emph{J. ACM}, 43\penalty0 (3):\penalty0 431–473, may 1996.
\newblock ISSN 0004-5411.
\newblock \doi{10.1145/233551.233553}.
\newblock URL \url{https://doi.org/10.1145/233551.233553}.

\bibitem[Goodrich(2014)]{10.1145/2591796.2591830}
Michael~T. Goodrich.
\newblock Zig-zag sort: A simple deterministic data-oblivious sorting algorithm running in o(n log n) time.
\newblock In \emph{Proceedings of the Forty-Sixth Annual ACM Symposium on Theory of Computing}, STOC '14, page 684–693, New York, NY, USA, 2014. Association for Computing Machinery.
\newblock ISBN 9781450327107.
\newblock \doi{10.1145/2591796.2591830}.
\newblock URL \url{https://doi.org/10.1145/2591796.2591830}.

\bibitem[Goodrich et~al.(2011)Goodrich, Sitchinava, and Zhang]{10.1007/978-3-642-25591-5_39}
Michael~T. Goodrich, Nodari Sitchinava, and Qin Zhang.
\newblock Sorting, searching, and simulation in the mapreduce framework.
\newblock In Takao Asano, Shin-ichi Nakano, Yoshio Okamoto, and Osamu Watanabe, editors, \emph{Algorithms and Computation}, pages 374--383, Berlin, Heidelberg, 2011. Springer Berlin Heidelberg.
\newblock ISBN 978-3-642-25591-5.

\bibitem[Gribov et~al.(2017)Gribov, Vinayagamurthy, and Gorbunov]{Gribov2017StealthDBAS}
Alexey Gribov, Dhinakaran Vinayagamurthy, and Sergey Gorbunov.
\newblock Stealthdb: a scalable encrypted database with full sql query support.
\newblock \emph{Proceedings on Privacy Enhancing Technologies}, 2019:\penalty0 370 -- 388, 2017.
\newblock URL \url{https://api.semanticscholar.org/CorpusID:28591687}.

\bibitem[Han et~al.(2022)Han, Zhang, Feng, Liu, and Li]{DBLP:conf/icde/HanZFLL22}
Feng Han, Lan Zhang, Hanwen Feng, Weiran Liu, and Xiang{-}Yang Li.
\newblock Scape: Scalable collaborative analytics system on private database with malicious security.
\newblock In \emph{38th {IEEE} International Conference on Data Engineering, {ICDE} 2022, Kuala Lumpur, Malaysia, May 9-12, 2022}, pages 1740--1753. {IEEE}, 2022.
\newblock \doi{10.1109/ICDE53745.2022.00176}.
\newblock URL \url{https://doi.org/10.1109/ICDE53745.2022.00176}.

\bibitem[Hu(2021)]{10.1145/3452021.3458319}
Xiao Hu.
\newblock Cover or pack: New upper and lower bounds for massively parallel joins.
\newblock In \emph{Proceedings of the 40th ACM SIGMOD-SIGACT-SIGAI Symposium on Principles of Database Systems}, PODS'21, page 181–198, New York, NY, USA, 2021. Association for Computing Machinery.
\newblock ISBN 9781450383813.
\newblock \doi{10.1145/3452021.3458319}.
\newblock URL \url{https://doi.org/10.1145/3452021.3458319}.

\bibitem[Hu and Yi(2019)]{10.1145/3294052.3319698}
Xiao Hu and Ke~Yi.
\newblock Instance and output optimal parallel algorithms for acyclic joins.
\newblock In \emph{Proceedings of the 38th ACM SIGMOD-SIGACT-SIGAI Symposium on Principles of Database Systems}, PODS '19, page 450–463, New York, NY, USA, 2019. Association for Computing Machinery.
\newblock ISBN 9781450362276.
\newblock \doi{10.1145/3294052.3319698}.
\newblock URL \url{https://doi.org/10.1145/3294052.3319698}.

\bibitem[Hu and Yi(2020)]{10.1145/3375395.3387657}
Xiao Hu and Ke~Yi.
\newblock Parallel algorithms for sparse matrix multiplication and join-aggregate queries.
\newblock In \emph{Proceedings of the 39th ACM SIGMOD-SIGACT-SIGAI Symposium on Principles of Database Systems}, PODS'20, page 411–425, New York, NY, USA, 2020. Association for Computing Machinery.
\newblock ISBN 9781450371087.
\newblock \doi{10.1145/3375395.3387657}.
\newblock URL \url{https://doi.org/10.1145/3375395.3387657}.

\bibitem[Hu et~al.(2017)Hu, Tao, and Yi]{10.1145/3034786.3056110}
Xiao Hu, Yufei Tao, and Ke~Yi.
\newblock Output-optimal parallel algorithms for similarity joins.
\newblock In \emph{Proceedings of the 36th ACM SIGMOD-SIGACT-SIGAI Symposium on Principles of Database Systems}, PODS '17, page 79–90, New York, NY, USA, 2017. Association for Computing Machinery.
\newblock ISBN 9781450341981.
\newblock \doi{10.1145/3034786.3056110}.
\newblock URL \url{https://doi.org/10.1145/3034786.3056110}.

\bibitem[Islam et~al.(2012)Islam, Kuzu, and Kantarcioglu]{DBLP:conf/ndss/IslamKK12}
Mohammad~Saiful Islam, Mehmet Kuzu, and Murat Kantarcioglu.
\newblock Access pattern disclosure on searchable encryption: Ramification, attack and mitigation.
\newblock In \emph{19th Annual Network and Distributed System Security Symposium, {NDSS} 2012, San Diego, California, USA, February 5-8, 2012}. The Internet Society, 2012.
\newblock URL \url{https://www.ndss-symposium.org/ndss2012/access-pattern-disclosure-searchable-encryption-ramification-attack-and-mitigation}.

\bibitem[Ketsman and Suciu(2017)]{10.1145/3034786.3034788}
Bas Ketsman and Dan Suciu.
\newblock A worst-case optimal multi-round algorithm for parallel computation of conjunctive queries.
\newblock In \emph{Proceedings of the 36th ACM SIGMOD-SIGACT-SIGAI Symposium on Principles of Database Systems}, PODS '17, page 417–428, New York, NY, USA, 2017. Association for Computing Machinery.
\newblock ISBN 9781450341981.
\newblock \doi{10.1145/3034786.3034788}.
\newblock URL \url{https://doi.org/10.1145/3034786.3034788}.

\bibitem[Krastnikov et~al.(2020)Krastnikov, Kerschbaum, and Stebila]{10.14778/3407790.3407814}
Simeon Krastnikov, Florian Kerschbaum, and Douglas Stebila.
\newblock Efficient oblivious database joins.
\newblock \emph{Proc. VLDB Endow.}, 13\penalty0 (12):\penalty0 2132–2145, jul 2020.
\newblock ISSN 2150-8097.
\newblock \doi{10.14778/3407790.3407814}.
\newblock URL \url{https://doi.org/10.14778/3407790.3407814}.

\bibitem[Leighton(1984)]{10.1145/800057.808667}
Tom Leighton.
\newblock Tight bounds on the complexity of parallel sorting.
\newblock In \emph{Proceedings of the Sixteenth Annual ACM Symposium on Theory of Computing}, STOC '84, page 71–80, New York, NY, USA, 1984. Association for Computing Machinery.
\newblock ISBN 0897911334.
\newblock \doi{10.1145/800057.808667}.
\newblock URL \url{https://doi.org/10.1145/800057.808667}.

\bibitem[Leskovec and Krevl(2014)]{snapnets}
Jure Leskovec and Andrej Krevl.
\newblock {SNAP Datasets}: {Stanford} large network dataset collection.
\newblock \url{http://snap.stanford.edu/data}, June 2014.

\bibitem[Li et~al.(2023{\natexlab{a}})Li, Zhao, Chen, Tan, Li, Wang, Mi, Xia, Li, and Chen]{288576}
Mingyu Li, Xuyang Zhao, Le~Chen, Cheng Tan, Huorong Li, Sheng Wang, Zeyu Mi, Yubin Xia, Feifei Li, and Haibo Chen.
\newblock Encrypted databases made secure yet maintainable.
\newblock In \emph{17th USENIX Symposium on Operating Systems Design and Implementation (OSDI 23)}, pages 117--133, Boston, MA, July 2023{\natexlab{a}}. USENIX Association.
\newblock ISBN 978-1-939133-34-2.
\newblock URL \url{https://www.usenix.org/conference/osdi23/presentation/li-mingyu}.

\bibitem[Li et~al.(2023{\natexlab{b}})Li, Li, and Gao]{10.14778/3583140.3583158}
Xiang Li, Fabing Li, and Mingyu Gao.
\newblock Flare: A fast, secure, and memory-efficient distributed analytics framework.
\newblock \emph{Proc. VLDB Endow.}, 16\penalty0 (6):\penalty0 1439–1452, feb 2023{\natexlab{b}}.
\newblock ISSN 2150-8097.
\newblock \doi{10.14778/3583140.3583158}.
\newblock URL \url{https://doi.org/10.14778/3583140.3583158}.

\bibitem[Li et~al.(2023{\natexlab{c}})Li, Sun, Luo, and Gao]{soda}
Xiang Li, Nuozhou Sun, Yunqian Luo, and Mingyu Gao.
\newblock Soda: A set of fast oblivious algorithms in distributed secure data analytics.
\newblock \emph{Proc. VLDB Endow.}, 16\penalty0 (7):\penalty0 1671–1684, mar 2023{\natexlab{c}}.
\newblock ISSN 2150-8097.
\newblock \doi{10.14778/3587136.3587142}.
\newblock URL \url{https://doi.org/10.14778/3587136.3587142}.

\bibitem[Liagouris et~al.(2023)Liagouris, Kalavri, Faisal, and Varia]{secrecy}
John Liagouris, Vasiliki Kalavri, Muhammad Faisal, and Mayank Varia.
\newblock {SECRECY:} secure collaborative analytics in untrusted clouds.
\newblock In Mahesh Balakrishnan and Manya Ghobadi, editors, \emph{20th {USENIX} Symposium on Networked Systems Design and Implementation, {NSDI} 2023, Boston, MA, April 17-19, 2023}, pages 1031--1056. {USENIX} Association, 2023.

\bibitem[Mitzenmacher and Upfal(2005)]{10.5555/1076315}
Michael Mitzenmacher and Eli Upfal.
\newblock \emph{Probability and Computing: Randomized Algorithms and Probabilistic Analysis}.
\newblock Cambridge University Press, USA, 2005.
\newblock ISBN 0521835402.

\bibitem[Mohassel et~al.(2020)Mohassel, Rindal, and Rosulek]{fastdatabasejoin}
Payman Mohassel, Peter Rindal, and Mike Rosulek.
\newblock Fast database joins and psi for secret shared data.
\newblock In \emph{Proceedings of the 2020 ACM SIGSAC Conference on Computer and Communications Security}, page 1271–1287. Association for Computing Machinery, 2020.

\bibitem[Ngai et~al.(2024)Ngai, Demertzis, Chamani, and Papadopoulos]{sp24}
N.~Ngai, I.~Demertzis, J.~Ghareh Chamani, and D.~Papadopoulos.
\newblock Distributed \& scalable oblivious sorting and shuffling.
\newblock In \emph{2024 IEEE Symposium on Security and Privacy (SP)}, pages 156--156, Los Alamitos, CA, USA, may 2024. IEEE Computer Society.
\newblock \doi{10.1109/SP54263.2024.00153}.
\newblock URL \url{https://doi.ieeecomputersociety.org/10.1109/SP54263.2024.00153}.

\bibitem[Ohrimenko et~al.(2015)Ohrimenko, Costa, Fournet, Gkantsidis, Kohlweiss, and Sharma]{10.1145/2810103.2813695}
Olga Ohrimenko, Manuel Costa, C\'{e}dric Fournet, Christos Gkantsidis, Markulf Kohlweiss, and Divya Sharma.
\newblock Observing and preventing leakage in mapreduce.
\newblock In \emph{Proceedings of the 22nd ACM SIGSAC Conference on Computer and Communications Security}, CCS '15, page 1570–1581, New York, NY, USA, 2015. Association for Computing Machinery.
\newblock ISBN 9781450338325.
\newblock \doi{10.1145/2810103.2813695}.
\newblock URL \url{https://doi.org/10.1145/2810103.2813695}.

\bibitem[Ohrimenko et~al.(2016)Ohrimenko, Schuster, Fournet, Mehta, Nowozin, Vaswani, and Costa]{10.5555/3241094.3241143}
Olga Ohrimenko, Felix Schuster, C\'{e}dric Fournet, Aastha Mehta, Sebastian Nowozin, Kapil Vaswani, and Manuel Costa.
\newblock Oblivious multi-party machine learning on trusted processors.
\newblock SEC'16, page 619–636, USA, 2016. USENIX Association.
\newblock ISBN 9781931971324.

\bibitem[Poddar et~al.(2021)Poddar, Kalra, Yanai, Deng, Popa, and Hellerstein]{263796}
Rishabh Poddar, Sukrit Kalra, Avishay Yanai, Ryan Deng, Raluca~Ada Popa, and Joseph~M. Hellerstein.
\newblock Senate: A {Maliciously-Secure} {MPC} platform for collaborative analytics.
\newblock In \emph{30th USENIX Security Symposium (USENIX Security 21)}, pages 2129--2146. USENIX Association, August 2021.
\newblock ISBN 978-1-939133-24-3.
\newblock URL \url{https://www.usenix.org/conference/usenixsecurity21/presentation/poddar}.

\bibitem[Popa et~al.(2011)Popa, Redfield, Zeldovich, and Balakrishnan]{10.1145/2043556.2043566}
Raluca~Ada Popa, Catherine M.~S. Redfield, Nickolai Zeldovich, and Hari Balakrishnan.
\newblock Cryptdb: Protecting confidentiality with encrypted query processing.
\newblock In \emph{Proceedings of the Twenty-Third ACM Symposium on Operating Systems Principles}, SOSP '11, page 85–100, New York, NY, USA, 2011. Association for Computing Machinery.
\newblock ISBN 9781450309776.
\newblock \doi{10.1145/2043556.2043566}.
\newblock URL \url{https://doi.org/10.1145/2043556.2043566}.

\bibitem[Priebe et~al.(2018)Priebe, Vaswani, and Costa]{8418608}
Christian Priebe, Kapil Vaswani, and Manuel Costa.
\newblock Enclavedb: A secure database using sgx.
\newblock In \emph{2018 IEEE Symposium on Security and Privacy (SP)}, pages 264--278, 2018.
\newblock \doi{10.1109/SP.2018.00025}.

\bibitem[Ren et~al.(2022)Ren, Su, Gu, Wang, Li, Xie, Bian, Li, and Zhang]{10.14778/3574245.3574248}
Xuanle Ren, Le~Su, Zhen Gu, Sheng Wang, Feifei Li, Yuan Xie, Song Bian, Chao Li, and Fan Zhang.
\newblock Heda: Multi-attribute unbounded aggregation over homomorphically encrypted database.
\newblock \emph{Proc. VLDB Endow.}, 16\penalty0 (4):\penalty0 601–614, dec 2022.
\newblock ISSN 2150-8097.
\newblock \doi{10.14778/3574245.3574248}.
\newblock URL \url{https://doi.org/10.14778/3574245.3574248}.

\bibitem[Sasy et~al.(2022)Sasy, Johnson, and Goldberg]{10.1145/3548606.3560603}
Sajin Sasy, Aaron Johnson, and Ian Goldberg.
\newblock Fast fully oblivious compaction and shuffling.
\newblock In \emph{Proceedings of the 2022 ACM SIGSAC Conference on Computer and Communications Security}, CCS '22, page 2565–2579, New York, NY, USA, 2022. Association for Computing Machinery.
\newblock ISBN 9781450394505.
\newblock \doi{10.1145/3548606.3560603}.
\newblock URL \url{https://doi.org/10.1145/3548606.3560603}.

\bibitem[Services(2024)]{AWSEMR}
Amazon~Web Services.
\newblock Amazon emr pricing, 2024.
\newblock URL \url{https://aws.amazon.com/emr/pricing/}.
\newblock Accessed: 2024-08-12.

\bibitem[Stefanov et~al.(2018)Stefanov, Dijk, Shi, Chan, Fletcher, Ren, Yu, and Devadas]{10.1145/3177872}
Emil Stefanov, Marten~Van Dijk, Elaine Shi, T.-H.~Hubert Chan, Christopher Fletcher, Ling Ren, Xiangyao Yu, and Srinivas Devadas.
\newblock Path oram: An extremely simple oblivious ram protocol.
\newblock \emph{J. ACM}, 65\penalty0 (4), apr 2018.
\newblock ISSN 0004-5411.
\newblock \doi{10.1145/3177872}.
\newblock URL \url{https://doi.org/10.1145/3177872}.

\bibitem[Sun et~al.(2021)Sun, Wang, Li, and Li]{10.14778/3447689.3447705}
Yuanyuan Sun, Sheng Wang, Huorong Li, and Feifei Li.
\newblock Building enclave-native storage engines for practical encrypted databases.
\newblock \emph{Proc. VLDB Endow.}, 14\penalty0 (6):\penalty0 1019–1032, feb 2021.
\newblock ISSN 2150-8097.
\newblock \doi{10.14778/3447689.3447705}.
\newblock URL \url{https://doi.org/10.14778/3447689.3447705}.

\bibitem[Valiant(1990)]{10.1145/79173.79181}
Leslie~G. Valiant.
\newblock A bridging model for parallel computation.
\newblock \emph{Commun. ACM}, 33\penalty0 (8):\penalty0 103–111, aug 1990.
\newblock ISSN 0001-0782.
\newblock \doi{10.1145/79173.79181}.
\newblock URL \url{https://doi.org/10.1145/79173.79181}.

\bibitem[Wang et~al.(2022)Wang, Li, Li, Li, Tian, Su, Zhang, Ma, Yan, Sun, Cheng, Xie, and Zou]{10.14778/3554821.3554826}
Sheng Wang, Yiran Li, Huorong Li, Feifei Li, Chengjin Tian, Le~Su, Yanshan Zhang, Yubing Ma, Lie Yan, Yuanyuan Sun, Xuntao Cheng, Xiaolong Xie, and Yu~Zou.
\newblock Operon: An encrypted database for ownership-preserving data management.
\newblock \emph{Proc. VLDB Endow.}, 15\penalty0 (12):\penalty0 3332–3345, aug 2022.
\newblock ISSN 2150-8097.
\newblock \doi{10.14778/3554821.3554826}.
\newblock URL \url{https://doi.org/10.14778/3554821.3554826}.

\bibitem[Wang and Yi(2021)]{10.1145/3448016.3452808}
Yilei Wang and Ke~Yi.
\newblock Secure yannakakis: Join-aggregate queries over private data.
\newblock In \emph{Proceedings of the 2021 International Conference on Management of Data}, SIGMOD '21, page 1969–1981, New York, NY, USA, 2021. Association for Computing Machinery.
\newblock ISBN 9781450383431.
\newblock \doi{10.1145/3448016.3452808}.
\newblock URL \url{https://doi.org/10.1145/3448016.3452808}.

\bibitem[Wang and Yi(2022)]{10.1145/3517804.3524142}
Yilei Wang and Ke~Yi.
\newblock Query evaluation by circuits.
\newblock PODS '22, page 67–78, New York, NY, USA, 2022. Association for Computing Machinery.
\newblock ISBN 9781450392600.
\newblock \doi{10.1145/3517804.3524142}.
\newblock URL \url{https://doi.org/10.1145/3517804.3524142}.

\bibitem[Xu et~al.(2015)Xu, Cui, and Peinado]{7163052}
Yuanzhong Xu, Weidong Cui, and Marcus Peinado.
\newblock Controlled-channel attacks: Deterministic side channels for untrusted operating systems.
\newblock In \emph{Proceedings of the 2015 IEEE Symposium on Security and Privacy}, SP '15, page 640–656, USA, 2015. IEEE Computer Society.
\newblock ISBN 9781467369497.
\newblock \doi{10.1109/SP.2015.45}.
\newblock URL \url{https://doi.org/10.1109/SP.2015.45}.

\bibitem[YSU-Data-Lab(2024)]{TPCHSKEW}
YSU-Data-Lab.
\newblock Tpc-h-skew, 2024.
\newblock URL \url{https://github.com/YSU-Data-Lab/TPC-H-Skew}.
\newblock Accessed: 2024-08-12.

\bibitem[Zaharia et~al.(2012)Zaharia, Chowdhury, Das, Dave, Ma, McCauley, Franklin, Shenker, and Stoica]{10.5555/2228298.2228301}
Matei Zaharia, Mosharaf Chowdhury, Tathagata Das, Ankur Dave, Justin Ma, Murphy McCauley, Michael~J. Franklin, Scott Shenker, and Ion Stoica.
\newblock Resilient distributed datasets: A fault-tolerant abstraction for in-memory cluster computing.
\newblock In \emph{Proceedings of the 9th USENIX Conference on Networked Systems Design and Implementation}, NSDI'12, page~2, USA, 2012. USENIX Association.

\bibitem[Zheng et~al.(2017)Zheng, Dave, Beekman, Popa, Gonzalez, and Stoica]{opaque}
Wenting Zheng, Ankur Dave, Jethro~G. Beekman, Raluca~Ada Popa, Joseph~E. Gonzalez, and Ion Stoica.
\newblock Opaque: An oblivious and encrypted distributed analytics platform.
\newblock In \emph{Proceedings of the 14th USENIX Conference on Networked Systems Design and Implementation}, NSDI'17, page 283–298, USA, 2017. USENIX Association.
\newblock ISBN 9781931971379.

\bibitem[Zhu et~al.(2021)Zhu, Cheng, Liu, and Guo]{10.14778/3476311.3476351}
Jinwei Zhu, Kun Cheng, Jiayang Liu, and Liang Guo.
\newblock Full encryption: An end to end encryption mechanism in gaussdb.
\newblock \emph{Proc. VLDB Endow.}, 14\penalty0 (12):\penalty0 2811–2814, jul 2021.
\newblock ISSN 2150-8097.
\newblock \doi{10.14778/3476311.3476351}.
\newblock URL \url{https://doi.org/10.14778/3476311.3476351}.

\bibitem[Zhuang et~al.(2004)Zhuang, Zhang, and Pande]{10.1145/1037947.1024403}
Xiaotong Zhuang, Tao Zhang, and Santosh Pande.
\newblock Hide: an infrastructure for efficiently protecting information leakage on the address bus.
\newblock \emph{SIGARCH Comput. Archit. News}, 32\penalty0 (5):\penalty0 72–84, oct 2004.
\newblock ISSN 0163-5964.
\newblock \doi{10.1145/1037947.1024403}.
\newblock URL \url{https://doi.org/10.1145/1037947.1024403}.

\end{thebibliography}

\end{document}